\documentclass{iopart}

\usepackage{iopams}
\usepackage{graphicx}

\newcommand{\x}{\bi{r}}
\newcommand{\kk}{\bi{k}}
\newcommand{\qq}{\bi{q}}
\newcommand{\aver}[1]{\left\langle #1 \right\rangle}

\newcommand{\tensor}[1]{\mathsf{#1}}

\begin{document}

\title{Radiative energy transfer in disordered photonic crystals}

\author{M. V. Erementchouk$^{1}$\footnote{The present address: NanoScience Technology Center, University of Central
Florida, Orlando, FL 32826}, %
L. I. Deych$^{2}$, 
H. Noh$^{1}$, 
H. Cao$^{1}$\footnote{The present address: Dept. of Applied Physics, Yale University, New Haven, CT 06520}, 
A.~A.~Lisyansky$^{2}$}

\address{$^{1}$Department of Physics and Astronomy, Northwestern
University, Evanston, IL 60208, \\
$^{2}$Physics Department, Queens College, City
University of New York, Flushing, New York 11367}
\ead{merement@mail.ucf.edu}

\begin{abstract}
The difficulty of description of the radiative transfer in disordered
photonic crystals arises from the necessity to consider on the equal footing
the wave scattering by periodic modulations of the dielectric function and
by its random inhomogeneities. We resolve this difficulty by approaching
this problem from the standpoint of the general multiple scattering theory
in media with arbitrary regular profile of the dielectric function. We use
the general asymptotic solution of the Bethe-Salpeter equation in order to
show that for a sufficiently weak disorder the diffusion limit in disordered
photonic crystals is presented by incoherent superpositions of the modes of
the ideal structure with weights inversely proportional to the respective
group velocities. The radiative transfer and the diffusion equations are
derived as a relaxation of long-scale deviations from this limiting
distribution. In particular, it is shown that in general the diffusion is
anisotropic unless the crystal has sufficiently rich symmetry, say, the
square lattice in 2D or the cubic lattice in 3D. In this case, the diffusion
is isotropic and only in this case the effect of the disorder can be
characterized by the single mean-free-path depending on frequency.
\end{abstract}


\maketitle

\section{Introduction}

The interest in structures with periodic modulations of the dielectric
function (photonic crystals \cite{JOANNOPOULOS:1995}) has been motivated
initially by the possibility to modify substantially the
spontaneous emission in such media. 
One of the main motives of the study, therefore, has been the existence of
the complete band gap \cite{HO:1990} when the propagation of light is
completely inhibited inside some frequency region. Only relatively recently
it has been realized that the periodicity of refractive index in photonic
crystals by itself results in a number of unusual properties even in absence
of the complete band gap
\cite{KOSAKA:1998,NOTOMI:2000,GUVEN:2004,ILIEW:2005,SHIN:2005,PELEG:2007}.
These properties do not necessarily require a strong contrast of the
periodic modulation, i.e. the ratio of the minimum and the maximum values of
the refractive index, and, particularly, can be observed even if the
contrast is much weaker than needed for the gap to open. One of the
illustrative examples is provided by dark modes
\cite{ROBERTSON:1992,SAKODA:1995,KARATHANOS:1998}, which are not coupled to
plane waves propagating outside the structure. The existence of these modes
is related to the point symmetry of the photonic crystal and, therefore, the
dark modes present even if the contrast is very low. The dark modes, as well
as other phenomena such as self-collimation and the negative refraction
demonstrate that periodic spatial modulations of the dielectric function
have more ways to affect the propagation of light than just producing a gap
in the spectrum.

Real photonic crystal structures always contain one or another type of
disorder regardless of a manufacturing procedure. It is crucially
important, therefore, 
to understand to which extent disorder affects properties of these
structures. This issue is of a great interest
because an interplay between periodic and random variations of the
refractive index creates new challenges for a theory of light propagation
in inhomogeneous media, and promises new and unusual effects in the
radiative transport.

The problem of the disorder in photonic crystals can be approached from two
perspectives. On one hand, there is an issue of effects of disorder on
spectral features of photonic crystals and their manifestation in such
characteristics as reflection or transmission spectra. This research
direction involves, for instance, studying such problems as a dependence of
the width of the photonic band gap on the degree of the disorder
\cite{SIGALAS:1999,LI:2000,KALITEEVSKI:2006}. A different type of questions
arise when one is concerned with effects of disorder on propagation of light
inside the photonic structure within the framework of the transport theory.
Problems considered in this case include diffusion in the photonic crystals
\cite{KOENDERINCK:2003,KOENDERINCK:2005} or enhanced backscattering
\cite{KOENDERINCK:2000,HUANG:2001,SIVACHENKO:2001}.

The main objective of the current paper is to develop a general theoretical
approach to wave transport in disordered photonic crystals, which would
systematically, from first principles, take into account the periodic nature
of the average refractive index. The microscopical nature of our approach
distinguishes it from earlier papers, which relied either on ad hoc
modifications of results obtained  within the multiple scattering theory in
statistically homogeneous media \cite{SIVACHENKO:2001,HUANG:2001}, or on
phenomenological assumptions regarding the distribution of the field in the
bulk of the photonic crystal \cite{KOENDERINCK:2005,KOENDERINCK:2000}. For
instance, it was assumed in the latter papers that propagation of light in
the bulk of the photonic structure is the same as in statistically
homogeneous random media, and the band structure of the photonic crystal
only manifests itself  within a narrow surface layer of the sample. This
idea is obviously based on the assumption that multiple scattering destroys
photonic modes of an ideal periodic structure so that the structure of the
electromagnetic field  in the bulk of the photonic crystal becomes
indistinguishable from that of a regular disordered medium. A number of
experimental and numerical results, however, cast doubts on this assumption.
For instance, it was shown in \cite{YAMILOV:2003} that photonic modes are
completely developed in samples with linear dimensions as small as just a
few periods. It was also determined experimentally that the mean-free-path,
$\ell$, [see Eqs.~(\ref{eq:mean_free_path_def}) and
(\ref{eq:single_ell_def})] due to disorder in real photonic crystals (see.
e.g. Table I in \cite{KOENDERINCK:2005_mfp}) substantially exceeds the
lattice constant of the photonic crystal. Even for relatively high
frequencies $\Omega a/2\pi c =1.6$ it was found that $\ell/a \approx 4$.
Thus, even in this least favourable case there are about hundred elementary
cells in a volume with the linear dimensions of the order of $\ell$.
Comparing these two results and invoking ideas of the separation of length
scales it is reasonable to expect that the underlying periodicity of the
photonic crystals must still manifest itself even in the presence of
disorder. In our paper we show that this expectation is indeed justified,
and demonstrate explicitly the effects of periodicity on wave transport in
the diffusive regime.

The standard multiple-scattering theory of wave transport in disordered
media depends significantly on the plane-wave representation of the
scattered field. The role of the plane waves is explicitly emphasized by the
use of the Wigner function in derivations of the radiative energy transfer
equations \cite{RYZHIK:1996,FERWERDA:1999,SHENDELEVA:2004,BAL:2006}. In
photonic crystals, however, approaches based on plane waves encounter
significant difficulties because plane waves are not normal modes of the
underlying ideal periodic structure. While the Wigner representation of the
field-field correlation function itself remains, of course, valid even in
such structures, the Wigner function, however, ceases to be a smooth
function of coordinates, which is essential for the derivation of the
radiative transfer and diffusion equations.

Qualitatively, one of the difficulties of the plane wave based approaches
is due to the fact that the plane waves are scattered not only by the
random fluctuations of the refractive index but also by its periodic
modulation. The latter is a purely deterministic process and is
responsible for the formation of photonic crystal modes, which can be
considered as coherent superpositions of the plane waves. Thus, in order
to describe wave transport in disordered photonic crystals one has to be
able to separate the deterministic contribution to the scattering from the
one caused by the disorder. This can be achieved by  developing the
transport theory on the base of photonic modes of an ideal crystal, but as
we will see even within this approach the discrimination between coherent
and incoherent processes is highly non-trivial. The second, more
technical, difficulty in adapting the standard multiple scattering theory
to photonic crystals arises from the fact that the theory developed for
statistically uniform media heavily depends on certain assumptions (e.g.
the translational invariance of the averaged Green's function) that  are
not valid in disordered photonic crystals.

In the present paper we resolve these difficulties and develop a consistent
multiple scattering theory of light transport in media with the
periodic-on-average dielectric function. Some of the results obtained are
actually valid also in media with arbitrary modulation of the background
(average) dielectric functions. We introduce the interpretation of the
field-field correlation function as the density matrix and show how it can
be used to separate coherent and incoherent contributions in the transport.
Using this idea we generalize the concept of specific intensities to the
case of photonic crystals and derive respective radiative transfer
equations. We also find an asymptotic solution of Bethe-Salpeter equation
describing a steady state intensity distribution in an infinite media far
away from sources, which we use to derive a diffusion equation in a steady
state regime describing long scale spatial relaxation of the intensity
toward the limiting distribution.

\section{Multiple scattering in disordered photonic crystals}
\label{sec:main_equations}

In the framework of the scalar model, the spatial distribution of the
wave field at frequency $\omega$ in a disordered photonic crystal is
governed by the Helmholtz equation
\begin{equation}\label{eq:Helmholtz_equation}
  \Delta E_\omega(\x) + \omega^2\widetilde{\epsilon}(\x)E_\omega(\x)
  = j(\x),
\end{equation}
where $j(\x)$ is the external source. Here and henceforth we use units with
$c=1$. In~(\ref{eq:Helmholtz_equation}) we have introduced the dielectric
function
\begin{equation}\label{eq:ref_index_separation}
 \widetilde{\epsilon}(\x) = \epsilon(\x) + \Delta\epsilon(\x),
\end{equation}
which consists of two components. The periodic part, $\epsilon(\x + \bi{a})
= \epsilon(\x)$, constitutes the photonic crystal with $\bi{a}$ being the
vector of lattice translations. The zero-mean random term,
$\Delta\epsilon(\x)$, describes the deviation of the dielectric function
from the ideal periodic form. We assume that the random part of the
refractive index is a zero-mean Gaussian random field, i.e. its statistical
properties are completely characterized by the covariance $K(\x_1,\x_2) =
\aver{\Delta\epsilon(\x_1)\Delta\epsilon(\x_2)}$.
In~\ref{sec:App_random_model} we provide a model of such inhomogeneities
relevant for disordered photonic crystals. With this
assumption~(\ref{eq:Helmholtz_equation}) can be readily analyzed by standard
diagrammatic techniques based on the Born series representation of the
solutions of the integral (Lippmann-Schwinger) formulation
of~(\ref{eq:Helmholtz_equation}), which were developed in the theory of
multiple wave scattering in statistically uniform media
\cite{Rytov_Principles_IV,VAN_ROSSUM:1998,Tsang_Advanced}. This approach is
virtually model independent and can be applied for structures with arbitrary
spatial profile of the deterministic part of the refractive index.


The Dyson equation for the Green's function of~(\ref{eq:Helmholtz_equation})
averaged over realizations of the disorder, $\bar{G}\equiv\aver{G}$, is
obtained using the standard diagrammatic technique and has the form
\begin{equation}\label{eq:gauss_Dyson_equation}
  \bar{G}_\omega(\x,\x') = G_0(\x,\x') + \int \rmd\x_1 \rmd\x_2\,
  G_0(\x,\x_1) \Sigma_\omega(\x_1,\x_2) \bar{G}_\omega(\x_2,\x'),
\end{equation}
where $\Sigma_\omega$ is the self-energy ``defined" as a sum of all
irreducible diagrams. We use somewhat cumbersome coordinate representation
in order to emphasize its generality and independence of a particular form
of the regular modulation $\epsilon(\x)$. The latter determines the
\textit{nonperturbed} Green's function, $G_0(\x,\x')$, which is assumed to
be known. In particular, in disordered photonic crystals $G_0(\x,\x')$ is
the Green's function of the ideal periodic structure.

The main difference between statistically uniform media and disordered
photonic crystals is reflected in this equation through symmetry properties
of the unperturbed Green's function. In periodic-on-average systems, this
function is invariant with respect to lattice translations and the group of
point symmetries of the underlying photonic structure, while in the uniform
disordered systems it has full translational and rotational symmetry.
In~\ref{sec:App_symmetries} we demonstrate that the averaged Green's
function and the self-energy posses the same translational and point
symmetries as $G_0(\x,\x')$. We utilize these properties by expanding all
related quantities in terms of photonic Bloch modes,
\begin{equation}\label{eq:photonic_Bloch}
 \Psi_{n,\kk}(\x) = \rme^{\rmi\kk \cdot \x} u_{\kk,n}(\x),
\end{equation}
where $u_{n,\kk}(\x)$ is periodic with the period of the photonic lattice,
$\kk$ is the Bloch wave vector lying inside the first Brillouin zone, and
$n$ enumerates photonic bands. Using these functions we introduce the matrix
representations of the quantities appearing
in~(\ref{eq:gauss_Dyson_equation}) according to
\begin{equation}\label{eq:Green_self-energy_expansion_bulk_PC}
\eqalign{
G_0(\x_1,\x_2) = & \sum_{\kk,n} g_n(\kk)
 \Psi_{\kk,n}(\x_1)\Psi^*_{\kk,n}(\x_2) , \cr
 \bar G_\omega(\x_1,\x_2) = & \sum_{\kk,n,m} G_{n,m}(\kk)
 \Psi_{\kk,n}(\x_1)\Psi^*_{\kk,m}(\x_2) , \cr
 \Sigma_\omega(\x_1,\x_2) = & \sum_{\kk,n,m}\Sigma_{n,m}(\kk)
 \epsilon(\x_1)\Psi_{\kk,n}(\x_1)\Psi^*_{\kk,m}(\x_2)\epsilon(\x_2).}
\end{equation}
The respective matrix elements in these expansions are
\begin{equation}\label{eq:aveGreen_self-energy_coefficients}
\eqalign{
 g_n(\kk) = \frac{1}{\omega^2 - \omega_n^2(\kk)}, \cr 
 G_{n,m}(\kk) = \frac{1}{\mathcal{V}^2}\int_\mathcal{V} \rmd\x_1\,\rmd\x_2\,
  \epsilon(\x_1)u^*_{\kk,n}(\x_1)\bar G_\omega(\x_1,\x_2)
  u_{\kk,m}(\x_2)\epsilon(\x_2), \cr
  \Sigma_{n,m}(\kk) = \frac{1}{\mathcal{V}^2}\int_\mathcal{V} \rmd\x_1\,\rmd\x_2\,
  u^*_{\kk,n}(\x_1)\Sigma_\omega(\x_1,\x_2)
  u_{\kk,m}(\x_2),}
\end{equation}
where $\omega_n^2(\kk)$ is the dispersion law of $n$-th band, the
integration is performed over the elementary cell of the ideal structure,
and $\mathcal{V}$ is the volume of the elementary cell.
Deriving~(\ref{eq:aveGreen_self-energy_coefficients}) we use the
orthogonality condition of the Bloch functions
\begin{equation}
 \int \rmd\x\, \epsilon(\x)\Psi_{n,\kk}^*(\x)\Psi_{m,\qq}(\x)
= \delta_{mn}\delta(\kk -\qq).
\end{equation}

Using~(\ref{eq:aveGreen_self-energy_coefficients}) we can rewrite
(\ref{eq:gauss_Dyson_equation}) in the matrix form
\begin{equation}\label{eq:Dyson_after_spectral_PC_bulk}
  G_{mn}(\kk) = g_{m}(\kk)\delta_{mn} + \sum_l
  g_{m}(\kk) \Sigma_{ml}(\kk)  G_{ln}(\kk).
\end{equation}
This equation emphasizes the fact that the translational invariance of the
self-energy prevents modes with different Bloch vectors to be mixed while
states corresponding to the same Bloch vector but belonging to different
bands are coupled by non-diagonal elements of the self-energy
$\Sigma_{ml}(\kk)$.

In order to analyze the general effect of the disorder, we separate the
diagonal, $\widehat{\Sigma}_{\mathrm{d}}(\kk)$, and the off-diagonal,
$\widehat{\Sigma}_{\mathrm{o}}(\kk)$, parts of the self-energy representing
the latter in the form
\begin{equation}\label{eq:self-energy_diagonal_and_off}
  \widehat{\Sigma}(\kk) = \widehat{\Sigma}_{\mathrm{d}}(\kk) + \widehat{\Sigma}_{\mathrm{o}}(\kk).
\end{equation}
The diagonal part $\widehat{\Sigma}_d(\kk)$ modifies each band
independently. It can be accounted for by introducing a modified Green's
function
\begin{equation}\label{eq:PC_diagonal_Green}
  \widetilde{G}_0(\kk) = \frac{1}{G_0^{-1}(\kk) - \widehat{\Sigma}_{\mathrm{d}}(\kk)},
\end{equation}
which is determined by the Dyson equation similar to the one written for the
standard case of a statistically homogeneous medium. In terms of the
modified Green's function~(\ref{eq:Dyson_after_spectral_PC_bulk}) takes the
form
\begin{equation}\label{eq:PC_Dyson_out_diagonal}
  \bar{G} = \widetilde{G}_0 + \widetilde{G}_0 \widehat{\Sigma}_{\mathrm{o}}
  \bar{G}.
\end{equation}
Eqs.~(\ref{eq:PC_diagonal_Green}) and
(\ref{eq:PC_Dyson_out_diagonal}) show the two-fold role of the
disorder in disordered photonic crystals. The disorder not only
modifies each band separately similar to the case of the
statistically homogeneous media, but also couples these modified
bands. It is important to note that the band coupling is a subject
of various selection rules. First, the most restrictive rule comes
from the translational symmetry of the self-energy. As has been
noted, it prevents states characterized by different Bloch
vectors to be coupled. In other words, from the perspective of a
band diagram one can have only ``vertical" coupling. The second rule
follows from the point symmetries. The self-energy transforms
according to identity representation of the symmetry group of a
given point in the reciprocal space of the photonic crystal. As a
result, its matrix elements between states corresponding to
different irreducible representations vanish. This selection rule
has important implications for high-symmetry points and directions.
In particular, it means that disorder does not lift the degeneracy
at the points where the degenerate states are described by different
or by multidimensional irreducible representations. The first can be
shown by direct calculations of the respective matrix elements of
the self-energy. The second follows from the following argument. The
states corresponding to a representation with a dimension higher
than $1$ can be coupled only with the states that transform
according to the same presentation. As a result the modified state
also transforms according to this presentation and, hence, the
respective state should also be degenerate. This implies, in
particular, that the degeneracy is not lifted at high-frequency
$\Gamma$-points.

In order to analyze the effect of the band coupling in more details, we
consider a two-band model when all bands but two (denoted by $1$ and $2$)
remain uncoupled. The solution of the Dyson equation describing the coupled
bands has the form
\begin{equation}\label{eq:PC_averaged_Green_two_band}
 \bar{G} = \frac{1}{\mathcal{D}_{12}}
 \left(
    \begin{array}{cc}
          g_2^{-1} - \Sigma_{22} & \Sigma_{12} \\
          \Sigma_{21} & g_1^{-1} - \Sigma_{11}
    \end{array}\right),
\end{equation}
where $g_i$ and $\Sigma_{ij}$ are the respective matrix elements of the
unperturbed Green's function and the self-energy, respectively. The zeros
of the function
\begin{equation}\label{eq:PC_D_function}
  \mathcal{D}_{12} = \left(g_1^{-1} - \Sigma_{11}\right)
  \left(g_2^{-1} - \Sigma_{22}\right) - \Sigma_{12}\Sigma_{21}
\end{equation}
give the spectrum of averaged excitations. Introducing
$\omega_{1,2}(\kk)$, the unperturbed dispersion laws of the interacting
bands, the poles of averaged Green's function can be written in the form
\begin{equation}\label{eq:PC_modified_bands}
  \bar{\omega}^2_i = \frac 1 2 \left(\widetilde{\omega}_1^2 + \widetilde{\omega}_2^2\right)
  \pm \frac{\Delta_{12}}{2}\sqrt{1+\eta_{12}}.
\end{equation}
Here $\widetilde{\omega}_{1,2}^2 = \omega_{1,2}^2 - \Sigma_{11,22}$
are the band frequencies after the diagonal modification,
$\Delta_{12} = \widetilde{\omega}_{1}^2 - \widetilde{\omega}_{2}^2$,
and
\begin{equation}\label{eq:band_coupling_parameter}
  \eta_{12} =
  4\frac{\Sigma_{12}\Sigma_{21}}{\Delta_{12}^2}
\end{equation}
is the parameter characterizing the strength of the band coupling. As one
could expect, this parameter is proportional to the matrix elements of the
self-energy between the coupled bands and is inversely proportional to the
frequency separation of the modified bands. When this parameter is small,
$\eta_{12}\ll 1$, the effect of the band coupling can be neglected and the
photonic crystals can be described in a single band approximation. It
should be noted, however, that in spite of formal similarity between the
averaged Green's function in the single band approximation and the
respective expression for the uniform disordered medium, the transport
properties of the two systems remain substantially different. The
condition   $\eta_{12}\ll 1$ can be easier satisfied at lower frequencies,
when the band separation is of the order of magnitude of the fundamental
band-gap. For higher frequency bands, however, the band coupling may play
a significant role even in the case of photonic crystals with weak
disorder and strong contrast of the refractive index.

\section{Transport in disordered photonic crystals}

\subsection{General formalism}

The transport properties of disordered media are characterized by the
field-field correlation function
\begin{equation}\label{eq:corr_fun_def}
 \rho_{\omega_1, \omega_2}(\x_1,\x_2) = \aver{E_{\omega_1}(\x_1)E^*_{\omega_2}(\x_2)},
\end{equation}
 which can be used to
describe transfer of energy and, generally, spatial distributions
and time evolutions of any quantity quadratic in the field. A
relation between $\rho_{\omega_1, \omega_2}(\x_1,\x_2)$ and the
external sources is provided by the intensity propagator
$\Pi(\x_1,\x_2;\x_1'\x_2')=\aver{G_{\omega_1}(\x_1,\x_1')G_{\omega_2}^*(\x_2,\x_2')}$
according to
\begin{equation}\label{eq:field_correlation_function}
  \rho_{\omega_1, \omega_2}(\x_1, \x_2) = \int \rmd\x_1'\rmd\x_2'\,
  \Pi(\x_1,\x_1';\x_2,\x_2')j_{\omega_1}(\x_1')j^*_{\omega_2}(\x_2').
\end{equation}
Using the standard diagrammatic technique one can show that the
intensity propagator satisfies the Bethe-Salpeter equation, which
can be written as
\begin{equation}\label{eq:gauss_Bethe_Salpeter}
\eqalign{\fl
  \Pi(\x_1,\x_2;\x_1'\x_2') = {} & \bar{G}_{\omega_1}(\x_1,\x_1')\bar{G^*}_{\omega_2}(\x_2,\x_2') + \cr
  &\int \rmd\x_3 \rmd\x_3' \rmd\x_4 \rmd\x_4'\,
  \bar{G}_{\omega_1}(\x_1, \x_3)\bar{G^*}_{\omega_2}(\x_2, \x_4)
  {U}_{\omega_1,\omega_2}(\x_3, \x_4; \x_3', \x_4') \cr
  & \qquad \qquad \times
    \Pi(\x_3',\x_4';\x_1'\x_2').}
\end{equation}

The kernel ${U}_{\omega_1,\omega_2}$ is the irreducible vertex presented
formally as a sum of irreducible diagrams \cite{Rytov_Principles_IV}. We
would like to emphasize that similarly to the Dyson equation, the
Bethe-Salpeter equation holds for an arbitrary regular spatial modulation of
the dielectric function, not necessarily periodic. We show
in~\ref{sec:App_symmetries} that regardless of the spatial dependence of the
regular part of the dielectric function, the irreducible vertex possesses an
important property of reciprocity
\begin{equation}\label{eq:irreducible_reciprocity}
  U_{\omega_1,\omega_2}(\x_1, \x_2; \x_1', \x_2') =
  U_{\omega_1,\omega_2}(\x_1', \x_2'; \x_1, \x_2).
\end{equation}
Additionally, in disordered photonic crystals this quantity is invariant
with respect to lattice translations (see~\ref{sec:App_symmetries})
suggesting the following representation for the vortex
\begin{equation}\label{eq:irreducible_Bloch}
\eqalign{\fl
 U_{\omega_1,\omega_2}(\x_1, \x_2; \x_1', \x_2') = {} & \sum_{v_{1,2}, l_{1,2}, \qq_i}
 \epsilon(\x_1)\epsilon(\x_2) \epsilon(\x_1')\epsilon(\x_2') \cr
 & \qquad U^{v_1,v_2}_{l_1, l_2}(\qq_1, \qq_2; \qq_3, \qq_4) 
 \delta\left(\overline{\qq_1 + \qq_4 - \qq_2 - \qq_3}\right) \cr
 & \qquad  \Psi_{\qq_1,v_1}(\x_1)\Psi_{\qq_4,l_2}(\x_2')
 \Psi^*_{\qq_2,v_2}(\x_2)\Psi^*_{\qq_3,l_1}(\x_1'),
}
\end{equation}
where the bar over a vector denotes the vector reduced to the first
Brillouin zone.

Using Bethe-Salpeter equation, Eqs.~(\ref{eq:gauss_Bethe_Salpeter}) and
(\ref{eq:field_correlation_function}) one can derive an equation for the
field-field correlation function  $\rho_{\omega_1, \omega_2}(\x_1,\x_2)$. We
present it in an integro-differential form, which is the most convenient for
further analysis. To derive such an equation one can apply operators
\begin{equation}\label{eq:coherent_operators}
  \frac{1}{\epsilon(\x_{1,2})}\Delta_{1,2} + \omega^2_{1,2},
\end{equation}
where indices $1$ and $2$ indicate a coordinate acted upon by the Laplacian,
to both sides of~(\ref{eq:field_correlation_function}).
As a result, in the region of space free of external sources, one
has
\begin{equation}\label{eq:rho_tansport_equation}
\eqalign{\fl
  \left[\frac{1}{\epsilon(\x_1)}\Delta_1 - \frac{1}{\epsilon(\x_2)}\Delta_2 + \omega_1^2 - \omega_2^2\right] &
  \rho_{\omega_1, \omega_2}(\x_1, \x_2) = \cr
  & \int d\x_1'd\x_2'\,\mathcal{F}_{\omega_1, \omega_2}(\x_1, \x_2; \x_1',
  \x_2')\rho_{\omega_1, \omega_2}(\x_1', \x_2'),}
\end{equation}
where
\begin{equation}\label{eq:scattering_kernel}
\eqalign{\fl
  \mathcal{F}_{\omega_1, \omega_2}(\x_1, \x_2; \x_1', \x_2') & = \frac{1}{\epsilon(\x_1)}\Sigma_{\omega_1}(\x_1,
  \x_1')\delta(\x_2 - \x_2') - \frac{1}{\epsilon(\x_2)}\Sigma_{\omega_2}^*(\x_2, \x_2')\delta(\x_1 - \x_1')\cr
  + & \int \rmd\x_1'' \rmd\x_2''
  \left[\frac{\delta(\x_1 - \x_1'')}{\epsilon(\x_1)}\bar{G}^*_{\omega_2}(\x_2, \x_2'')-
  \frac{\delta(\x_2 - \x_2'')}{\epsilon(\x_2)}
  \bar{G}_{\omega_1}(\x_1, \x_1'')\right] \cr
  & \qquad \times U_{\omega_1,\omega_2}(\x_1'',\x_2'';\x_1',\x_2').}
\end{equation}
This equation has been a subject of numerous investigations
\cite{Rytov_Principles_IV,LAGENDIJK:1996}, most of which were concerned with
the radiative transfer or diffusion regimes in statistically homogeneous
media. In many of those works, diffusion was understood as a transport
process characterized by asymptotically slow (both in time and space)
changes of the field intensity. In the spectral domain this behaviour
manifests itself in the form of the characteristic ``diffusion" pole of the
intensity propagator, which is proportional to $\left(i\Omega -
DQ^2\right)^{-1}$ in the limits $\Omega\rightarrow 0$, $Q\rightarrow 0$.
Here frequency and wave vector transfers, $\Omega=\omega_1-\omega_2$ and
$Q=q_1-q_2$, characterize slow spatiotemporal dynamics of intensity, and
wave vectors $\bi q_1$ and $\bi q_2$ arise from the plane-wave
representation of the scattered field. This reliance on the plane waves
significantly complicates generalization of standard microscopic derivations
of radiative transfer or diffusive equations to the case of disordered
photonic crystals.

In order to better recognize the source of these difficulties and find a way
to circumvent them, it is necessary to re-examine basic physical ideas about
diffusion of light in disordered media. As a first step in this direction,
in the current paper we consider time independent spatial distribution of
the wave intensity in an infinite medium far away from the sources. This
regime arises in the case of a monochromatic source, when the field in the
structure harmonically depends on time, $\propto \exp(-i\omega t)$, so that
$\Omega=0$. In this case, the equation for the field-field correlation
function $\rho = \aver{E_\omega E^*_\omega}$ (hereafter we omit the lower
index corresponding to frequency) can be obtained
from~(\ref{eq:rho_tansport_equation}) by setting $\omega_1 = \omega_2 =
\omega$:
\begin{equation}\label{eq:rho_tansport_equation_steady}
\fl  \left[\frac{1}{\epsilon(\x_1)}\Delta_1 -
  \frac{1}{\epsilon(\x_2)}\Delta_2\right]
  \rho(\x_1, \x_2) =
  \int \rmd\x_1'\rmd\x_2'\,\mathcal{F}(\x_1, \x_2; \x_1',
  \x_2')\rho(\x_1', \x_2'),
\end{equation}
where $\mathcal{F} \equiv \mathcal{F}_{\omega, \omega}$.
Using~(\ref{eq:rho_tansport_equation_steady}), the optical theorem (the Ward
identity) can be derived \cite{Rytov_Principles_IV} in the form
$\mathcal{F}(\x, \x; \x_1', \x_2') \equiv 0$, which is especially useful
from the technical point of view. Integrating this equation over $\x$ we
obtain the Ward identity in the form
\begin{equation}\label{eq:WI_kernel}
 \fl\Sigma(\x_2,\x_1) - \Sigma^*(\x_1,\x_2)
  =\int \rmd\x_1''\rmd\x_2''
  \left[\bar{G}(\x_2'',\x_1'') - \bar{G}^*(\x_1'',\x_2'')\right] U(\x_1'',\x_2''; \x_1, \x_2).
\end{equation}

The physical picture of the transport in disordered media is developed using
close relation of the function $\rho(\x_1, \x_2)$ to both transport
characteristics and  coherence properties of the field. The description of
transport (e. g. the energy density and the flux) is based on the property
that the averaged values of any quantity quadratic in field can be expressed
in terms of convolution of respective operators with $\rho(\x_1, \x_2)$ [see
Eqs.~(\ref{eq:density_energy_trace}) and
(\ref{eq:average_Poynting_incoherent}) below]. The coherence properties are
described considering $\rho(\x_1, \x_2)$ as the coherence function
\cite{MANDEL:1995,MANDEL:1965}.

A quantity, which simultaneously describes such transport related
characteristics as energy density and flux and coherence properties of the
system, is well known in quantum statistics. It is called the density matrix
\cite{FainIrreversibility}. Indeed, on the one hand, the density matrix can
be used to calculate current and energy densities, and on the other hand,
the density matrix describes mixed states, which can be characterized as
incoherent superpositions of pure or coherent states. Using the density
matrix analogy one can think of the field-field correlation function as a
characteristic of such a mixed state of the wave field in disordered media.
Separation of coherent and incoherent properties of the field would then
involve finding states whose incoherent superposition would reproduce the
field at the level of function $\rho(\x_1, \x_2)$. However, before
developing this idea any further, we need to demonstrate that the
field-field correlation function has indeed all the formal properties of the
density matrix.

We expand $\rho(\x_1,\x_2)$ in terms of the eigenstates of the nonperturbed
system, say, the modes of the ideal photonic crystal, writing
\begin{equation}\label{eq:density_state_representation}
  \rho(\x_1,\x_2) = \sum_{\mu,\nu} \rho_{\mu,\nu}\Psi_\mu(\x_1)
  \Psi_\nu^*(\x_2),
\end{equation}
where summation over indices $\mu$ and $\nu$ enumerating the eigenstates can
involve integration. It is easy to see from the definition of the
correlation function that coefficients $\rho_{\mu ,\nu}$ constitute a
Hermitian matrix. This matrix can be diagonalized, which corresponds to the
spectral representation of the statistical operator in quantum mechanics, by
means of a unitary transformation to another basis
\begin{equation}\label{eq:new_states}
  \bar{\Psi}_{\bar{\kappa}} = B_{\bar{\kappa}\mu}\Psi_\mu
\end{equation}
so that one has (Mercer's theorem \cite{MANDEL:1995,Riesz_FA})
\begin{equation}\label{eq:density_diagonalized}
\rho(\x_1,\x_2) =
\sum_{\bar{\kappa}}\rho_{\bar{\kappa}}\bar{\Psi}_{\bar{\kappa}}(\x_1)
  \bar{\Psi}^*_{\bar{\kappa}}(\x_2)
\end{equation}
with $\rho_{\bar{\kappa}} \geq 0$, which follows from the fact that
$\rho(\x_1,\x_2)$ is non-negatively defined \cite{WOLF:1982}. The diagonal
representation of the correlation function can be interpreted as an
\textit{incoherent} superposition or mixture of pure or coherent states
$\bar{\Psi}_{\bar{\kappa}}$, which, in turn, as follows
from~(\ref{eq:new_states}) are \textit{coherent} superpositions of the
states $\Psi_\mu$. In order to clarify the exact meaning of this expression
let us show that the correlation function can be used to calculate the
energy density of the field and its Poynting vector in much the same way as
the density matrix is used in quantum statistics.

The energy density of the field in a steady state can be presented
as
\begin{equation}\label{eq:energy_density}
  w(\x) = \frac{1}{2}\left[\omega^2\widetilde{\epsilon}(\x)|E(\x)|^2
  + |\nabla E(\x)|^2\right],
\end{equation}
where $\widetilde{\epsilon}(\x)$ is the total dielectric function
including both  regular and the random components. By introducing the
operator
\begin{equation}\label{eq:energy_operator_definition}
  \widehat{w} = -(\nabla_1 - \nabla_2)^2/4,
\end{equation}
one can show that the averaged energy density can be expressed in
terms of the function
\begin{equation}\label{eq:energy_generating}
  w(\x_1, \x_2) = \widehat{w}\rho(\x_1, \x_2)
\end{equation}
as
\begin{equation}\label{eq:average_energy_density}
  \aver{w(\bi{R})} = w(\bi{R}, \bi{R}),
\end{equation}
where we used the Helmholtz equation to arrive at~%
(\ref{eq:average_energy_density}). One can see that this equation
can be presented in the typical for quantum statistics form as
\begin{equation}\label{eq:density_energy_trace}
  \aver{w(\bi{R})} = \mathrm{Tr}\left[\widehat{\rho}\, \widehat{w}_\bi{R}\right],
\end{equation}
where $\widehat{w}_\bi{R} = \delta(\x_1 - \bi{R})\delta(\x_2 -
\bi{R})\widehat{w}$.

As an example let us consider the case when the density matrix has the form
of an incoherent superposition of the eigenstates $\Psi_\mu$ of the ideal
system. Then from~(\ref{eq:density_energy_trace}) we find
%
\begin{equation}\label{eq:density_energy_diagonal}
  \aver{w(\bi{R})} = \sum_{\mu} \rho_\mu w_\mu(\bi{R}),
\end{equation}
where we have introduced the energy density of the $\mu$-th mode
\begin{equation}\label{eq:mu_mode_energy_density}
 w_\mu(\bi{R}) = \widehat{w}\left.\Psi_\mu(\x_1)\Psi_\mu^*(\x_2)
 \right|_{\x_1=\x_2=\bi{R}}.
\end{equation}
This expression provides a clear explanation of the notion of incoherent
superposition. Indeed, the average energy of the field in this expression
is presented as a sum of \emph{energies} of individual modes
$\Psi_\mu(\x)$ as it is expected for the addition of incoherent fields as
oppose to the sum of the field amplitudes expected for the coherent
fields.

Similar expressions can be obtained for the average value of the Poynting
vector, $\bi{S} = \rmi\omega\left[E^*\nabla E - E\nabla E^*\right]/2$, which
can be calculated using the operator
\begin{equation}\label{eq:Poynting_operator}
  \widehat{\bi{S}} = -\frac{\rmi\omega}{2}\left(\nabla_1 -
  \nabla_2\right),
\end{equation}
which gives
\begin{equation}\label{eq:averaged_Poynting}
 \aver{\bi{S}(\bi{R})} = \bi{S}(\bi{R},
 \bi{R}),
\end{equation}
where
\begin{equation}\label{eq:flux_distribution}
  \bi{S}(\x_1,\x_2) =  \widehat{\bi{S}}\rho(\x_1,\x_2).
\end{equation}
In the case when the density matrix is diagonal in the basis of the
eigenfunctions of the ideal system we see again that the average Poynting
vector is a sum of Poynting vectors of each mode, which indicates the
absence of any interference effects in the superposition of modes
$\Psi_\mu(\x)$:
\begin{equation}\label{eq:average_Poynting_incoherent}
 \aver{\bi{S}(\bi{R})} = \mathrm{Tr}\left[\widehat{\rho}\, \widehat{\bi{S}}_\bi{R}\right]
 = \sum_{\mu} \rho_\mu \bi{S}_\mu(\bi{R}),
\end{equation}
where $\widehat{\bi{S}}_\bi{R} = \delta(\x_1 - \bi{R})\delta(\x_2 -
\bi{R})\widehat{\bi{S}}$ and $\bi{S}_\mu(\bi{R})$ is the distribution of the
Poynting vector in the $\mu$-th mode
\begin{equation}\label{eq:mu_mode_Poynting_vector}
  \bi{S}_\mu(\bi{R}) = \widehat{\bi{S}}\left.\Psi_\mu(\x_1)\Psi_\mu^*(\x_2)
 \right|_{\x_1=\x_2=\bi{R}}.
\end{equation}

It is seen from Eqs.~(\ref{eq:density_energy_diagonal}) and
(\ref{eq:average_Poynting_incoherent}) that $\rho_\mu$ have the meaning of
the weights of the incoherent superposition. More generally, if we normalize
$\rho(\x_1,\x_2)$ in such a way that $\mathrm{Tr}[\widehat{\rho}]\equiv1$,
we can interpret $\rho_{\bar{\kappa}}$, the eigenvalues of the matrix
$\rho_{\mu,\nu}$, as the distribution function in the space of the states
$\bar{\Psi}_{\bar{\kappa}}$. The values of any quantity, quadratic in field
averaged over the disorder realizations, therefore, can be calculated as the
average over the distribution function $\rho_{\bar{\kappa}}$ in the similar
way as it is done in quantum
mechanics. We would like to emphasize here 
that the emergence of the incoherent superposition in the problem of wave
propagation is directly related to averaging over realizations of disorder.
Without averaging the density matrix defined
in~(\ref{eq:density_diagonalized}) would have a form $\rho_{\mu,\nu}\propto
a_{\mu}a_{\nu}$, and matrices of such form have a single non-zero
eigenvalue. As a result, the sums in~(\ref{eq:density_energy_diagonal}) and
(\ref{eq:average_Poynting_incoherent}) would consist of just one term
indicating that $\rho$ represents a pure or coherent state.

The notion of incoherent superposition expressed by
Eqs.~(\ref{eq:density_energy_diagonal}) and
(\ref{eq:average_Poynting_incoherent}) also allows one to provide a
physical meaning for the states $\bar{\Psi}_{\bar{\kappa}}$ diagonalizing
the density matrix. The spatial field distribution in a random media can
be in principle presented as a linear combination of functions from any
full system: plane waves, Bloch waves, etc. The concept of normal modes as
well defined spatial distributions of the fields that can be excited
separately one from another is not very useful here. Indeed the
distribution of the field in a random media is so complex that at any
given frequency it is impossible to excite a single mode out of infinitely
many degenerate modes. In this situation, functions
$\bar{\Psi}_{\bar{\kappa}}$ play a special role as such distributions of
the field whose linear combination is purely incoherent in the sense of
Eqs.~(\ref{eq:density_energy_diagonal}) and
(\ref{eq:average_Poynting_incoherent}). The form of these functions is
determined by remaining coherence effects in the scattered waves, and
thus, by using the density matrix formalism we achieve a separation
between coherent and incoherent contributions to the energy transport in
disordered systems. In other words, one can say that the functions
diagonalizing the density matrix describe the modes, which provide the
energy transfer.

In order to illustrate these general idea, let us consider a case of wave
scattering in a homogeneous random medium. In the case, of an infinite
medium and asymptotically far away from the sources the field-field
correlation function restores its translational invariance:
$\rho(\x_1,\x_2)\rightarrow \rho(\x_1-\x_2)$. Using plane waves as a basis
we can rewrite~(\ref{eq:density_state_representation}) for this particular
case as
\begin{equation}\label{eq:homog_medium_example}
\rho(\x_1,\x_2)=\int \rmd\qq \rmd\kk \, \rho_{\mathrm{un}}(\kk)\delta(\kk-\qq).
\end{equation}
One can see that the density matrix in this case is diagonal in the basis of
the plane waves so that they are responsible for the incoherent transport.

\subsection{The field-field correlation in an infinite photonic crystal: asymptotic behaviour}

In this subsection we consider a solution of the steady-state Bethe-Salpeter
equation (\ref{eq:rho_tansport_equation_steady}) for the correlation
function $\rho(\x_1,\x_2)$ valid in an infinite photonic crystal
asymptotically far away from the sources. Besides providing an important
non-trivial example of the application of our formalism, this solution shows
how the periodicity of the underlying photonic structure affects asymptote
of the intensity distribution in the disordered structure and provides a
starting point for deriving the diffusion equation. Using
Eqs.~(\ref{eq:irreducible_reciprocity}) and (\ref{eq:WI_kernel}) one can
check that~(\ref{eq:rho_tansport_equation_steady}) is solved by
\begin{equation}\label{eq:density_limit}
  \rho^{(\infty)}(\x_1, \x_2) = \frac{1}{2\,\rmi N}
  \left[\bar G^*(\x_2,\x_1) - \bar G(\x_1,\x_2)\right],
\end{equation}
where $N$ is a normalization constant independent of coordinates. Using the
reciprocity theorem \cite{Rytov_Principles_IV} this is rewritten as
\begin{equation}\label{eq:reciprocity}
\rho^{(\infty)}(\x_1, \x_2) = -\mathrm{Im}\left[ \bar G(\x_1, \x_2)\right]/N.
\end{equation}
This solution is valid in a general case regardless of the specific form of
the regular modulation of the dielectric function and the distribution of
the disorder. In the case of statistically uniform media it is reduced to a
form found in Refs.~\cite{BARABANENKOV:1991} and \cite{APALKOV:2004}. Thus
we can consider this function as the asymptotic distribution of the
field-field correlator, which can be called for this reason an equilibrium
density matrix.

In the case of disordered photonic crystals, function $\rho^{(\infty)}(\x_1,
\x_2)$ can be expanded in terms of normal modes of the underlying periodic
structure. According to~(\ref{eq:Green_self-energy_expansion_bulk_PC}) we
can present this expansion in the following form
\begin{equation}\label{eq:equilibrium_PC_basis}
 \rho^{(\infty)}(\x_1, \x_2) = \sum_{\kk,n,m}
 \rho^{(\infty)}_{m,n}(\kk)\Psi_{\kk,n}(\x_1)\Psi^*_{\kk,m}(\x_2),
\end{equation}
where
\begin{equation}\label{eq:density_matrix_equil}
\eqalign{
\rho^{(\infty)}_{m,n}(\kk) & = \left[\bar G_{n,m}^*(\kk) - \bar
G_{m,n}(\kk)\right]/2\,\rmi N \cr
 & = \frac{1}{N\mathcal{V}^2}\int_\mathcal{V}
\rmd\x_1\,\rmd\x_2\,
  \epsilon(\x_1)u^*_{\kk,n}(\x_1) \mathrm{Im}\,\bar G_\omega(\x_1,\x_2)
  u_{\kk,m}(\x_2)\epsilon(\x_2).}
\end{equation}
This expression shows that in the basis of normal modes of an ideal periodic
structure, the density matrix is diagonal with respect to the quasi-wave
vector $\kk$, but is not diagonal with respect to the band indices.
Following Eqs.~(\ref{eq:density_state_representation}),
(\ref{eq:new_states}), and (\ref{eq:density_diagonalized}) we diagonalized
the matrix $\rho^{(\infty)}_{m,n}(\kk)$ by a unitary transformation
\begin{equation}\label{eq:PC_new_basis}
 \bar{\Psi}_{\kk,\bar{m}} = \sum_n B_{\bar{m}n}(\kk)\Psi_{\kk,n}.
\end{equation}
Since the diagonalization procedure involves only band indices and leaves
the Bloch vector intact, functions $\bar{\Psi}_{\kk,\bar{m}}(\x)$ can also
be presented in the Bloch form similar to~(\ref{eq:photonic_Bloch})
\begin{equation}\label{eq:new_states_Bloch_form}
  \bar{\Psi}_{\kk, \bar{m}}(\x) = \rme^{\rmi\kk \cdot \x} \bar{u}_{\kk,
  \bar{m}}(\x).
\end{equation}
Transformation (\ref{eq:PC_new_basis}) preserves the scalar product
because of unitarity so that
\begin{equation}\label{eq:new_states_orthogonality}
  \int \rmd\x\, \epsilon(\x) \bar{\Psi}_{\kk,\bar{m}}^*(\x)
  \bar{\Psi}_{\qq,\bar{n}}(\x) =
  \delta(\kk-\qq)\delta_{\bar{n}\bar{m}}.
\end{equation}
Using this property we normalize $\rho^{(\infty)}$ defining
\begin{equation}\label{eq:N_definition}
  N = -\pi \int_{\mathcal{V}}\rmd\x\, \epsilon(\x) \mathrm{Im}\,[\bar{G}(\x, \x)],
\end{equation}
where the integration is performed over the elementary cell of the ideal
structure. Equations~(\ref{eq:density_limit}) and (\ref{eq:N_definition})
define $\rho^{(\infty)}$ as the asymptotic form of density matrix, which
according to Eqs.~(\ref{eq:equilibrium_PC_basis}) and
(\ref{eq:PC_new_basis}) is an incoherent superposition of the states
$\bar{\Psi}_{\kk,\bar{m}}$ with respective weights.

As was discussed, the eigenvalues of $\rho^{(\infty)}$ define the
probability distribution function on the space of the states
$\bar{\Psi}_{\kk,\bar{m}}(\x)$ parametrized by the number of band $\bar{m}$
and by the point in the first Brillouin zone $\kk$. As follows from the
Dyson equation the singularities of the averaged Green's function and,
respectively, of the eigenvalues of $\rho^{(\infty)}_{m,n}$ are determined
by the dispersion law of the average excitations [see
e.g.~(\ref{eq:PC_D_function})]. For more detailed analysis we consider the
situation when we can neglect the band coupling so that the averaged Green's
function is given by~(\ref{eq:PC_diagonal_Green}). In this case, the
eigenvalues of the density matrix as functions of the quasi-wave vector have
the largest values, when $\kk$ obeys the dispersion equation for a given
frequency. In other words, these eigenvalues reach maximum values on
equifrequency surfaces, $F_m(\omega)$, corresponding to different bands of
the ideal photonic crystal. The width of these maxima is proportional to
$\mathrm{Im}[{\Sigma_d}_m(\kk)]$
[see~(\ref{eq:self-energy_diagonal_and_off})]. If the disorder is weak in
the sense of the Ioffe-Regel criterion (see below), then for frequencies,
which do not lie at band edges, we can obtain
\begin{equation}\label{eq:density_equilibrium_group_off}
\fl \rho^{(\infty)}(\x_1, \x_2) \approx \frac{\pi}{2N\omega} \sum_{m}
 \int_{F_m(\omega)}\rmd\kk\,
 \frac{1}{|v_m(\kk)|}\rme^{-|\boldsymbol{\gamma}_m(\kk) \cdot(\x_1 - \x_2)|}\Psi_{\kk,m}(\x_1)\Psi^*_{\kk,m}(\x_2),
\end{equation}
where the integration runs along the equifrequency surfaces, $\bi{v}_m(\kk)
= \nabla_\kk\omega_m(\kk)$ is the group velocity, and
\begin{equation}\label{eq:transverse_gamma_def}
  \boldsymbol{\gamma}_m(\kk) = -
  \frac{\mathrm{Im}[{\Sigma_{\mathrm{d}}}_m(\kk)]\bi{v}_m(\kk)}{2\,\omega
  \,v_m^2(\kk)} = \frac 1 2 \ell_m^{-1}(\kk) \widehat{\bi{v}}_m(\kk).
\end{equation}
Here $\widehat{\bi{v}}_m(\kk)$ is the unit vector along the direction of the
group velocity and we have expressed the imaginary part of the self-energy
in terms of the respective mean-free-path $\ell_m(\kk)$, which is defined
later in~(\ref{eq:mean_free_path_def}).

\begin{figure}
  \includegraphics[width=3in]{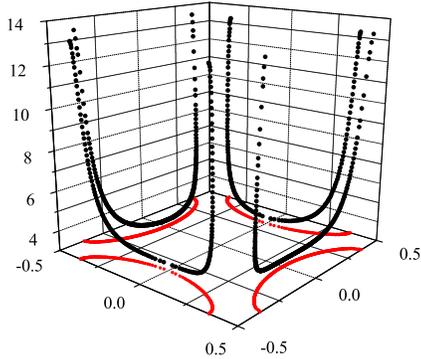}
  \caption{(Colour online) The dependence
  of $1/v_m(\kk)$ along the equifrequency surface corresponding to $\omega a/2\pi = 0.41$.
  The horizontal plane is the first Brillouin zone.
  Here and below the calculations are presented
 for a square lattice 2D
  PC made of dielectric cylinders with the contrast of the
  refractive indices $n=2$ and with the radius-period ratio $r/a =
  0.4$. }\label{fig:group_velocity}
\end{figure}

As follows from~(\ref{eq:density_equilibrium_group_off}) in the limit of
vanishing disorder, which corresponds to the on-shell approximation
\cite{Rytov_Principles_IV,LAGENDIJK:1996,VAN_TIGGELEN:1992,SOUKOULIS:1994,LIVDAN:1996}
in the standard theory of transport in statistically homogeneous media, the
density matrix takes the universal limit
\begin{equation}\label{eq:density_equilibrium_group}
 \rho^{(\infty)}_0(\x_1, \x_2) = \frac{\pi}{2N\omega} \sum_{m}
 \int_{F_m(\omega)}\rmd\kk\,
 \frac{1}{v_m(\kk)}\Psi_{\kk,m}(\x_1)\Psi^*_{\kk,m}(\x_2).
\end{equation}
The magnitude of the group velocity is, in general, not constant along the
equifrequency surfaces. As a result, different states are not equally
presented in the equilibrium distribution, as is seen from
Eqs.~(\ref{eq:density_equilibrium_group_off}) and
(\ref{eq:density_equilibrium_group}). In particular, if there are flat bands
\cite{SAKODA:2005} with low group velocity near the frequency $\omega$, then
these bands would give the main contribution to $\rho^{(\infty)}_0$. Another
example of a highly inhomogeneous distribution is provided by the
frequencies when an equifrequency surface touches the boundary of the
Brillouin zone. In this case, the magnitude of the group velocity becomes
very low at the points of contact resulting in an increased weight of the
respective states in the equilibrium distribution (see
Figure~\ref{fig:group_velocity}).

It should be noted that the on-shell approximation is poorly suited for
studying the correlation properties of the field. Indeed, the function
$\rho^{(\infty)}_0(\x_1, \x_2)$ does not tend to any limit as $|\x_1 - \x_2|
\to \infty$ while $\rho^{(\infty)}(\x_1, \x_2)$
in~(\ref{eq:density_equilibrium_group_off}) obviously vanishes in this
limit. Thus, for this purpose one has to use~(\ref{eq:density_limit}) or in
the case of a weak disorder the simpler
version~(\ref{eq:density_equilibrium_group_off}), as it has been practically
done in \cite{APALKOV:2004}, where additionally the approximation of
spherical equifrequency surfaces was used.

However, for studying transport properties the on-shell approximation may by
appropriate because, as follows from Eqs.~(\ref{eq:average_energy_density})
and (\ref{eq:averaged_Poynting}), the energy density and the Poynting vector
are determined by the behaviour of $\rho(\x_1, \x_2)$ near the diagonal
$\x_1 = \x_2$. We estimate the effect of the exponential term
in~(\ref{eq:density_equilibrium_group_off}) by calculating the average
energy density. The density matrix in the on-shell approximation yields the
energy density as a sum of contributions of different modes
[see~(\ref{eq:density_energy_diagonal})]
\begin{equation}\label{eq:energy_density_limit_PC}
  \aver{w(\bi{R})} = \sum_{\kk, m} \rho^{(\infty)}_{\kk, m} w_{\kk,
  m}(\bi{R}),
\end{equation}
where $w_{\kk, m}(\bi{R})$ is the energy density of the mode $\Psi_{\kk, m}$
of the ideal photonic crystal. The exponential term
in~(\ref{eq:density_equilibrium_group_off}) can be shown to modify each
contribution by the factor
\begin{equation}\label{eq:off_energy_modification}
  \sim 1 - \frac{1}{\left[4\omega \ell_m(\kk)\right]^2}.
\end{equation}
In order to evaluate the correction, we note that $\omega \ell_m(\kk)$ is a
parameter of the Ioffe-Regel type \cite{VAN_ROSSUM:1998}, which is expected
to be much larger than unity far from the localization regime, so that the
deviation of the equilibrium density matrix
from~(\ref{eq:density_equilibrium_group}) can be neglected in the analysis
of transport quantities.

The problem of special interest is what happens in an immediate vicinity of
the complete band gap. Formal application of~(\ref{eq:mean_free_path_def})
gives vanishing mean-free-path for the modes lying at the edge of the gap.
This means that approximation~(\ref{eq:density_equilibrium_group_off}) for
the density matrix is no longer valid and that the modes of the ideal
photonic crystal are not a good approximation for the modes
$\bar{\Psi}_{\bar{\mu}}$. Such nonperturbative reconstruction of the modes
constitutes a special problem and is not considered in the present paper.
Therefore, in what follows we restrict to the frequencies which are not too
close to the complete gap so that the condition $\omega \ell_m(\kk)>1$ can
always be fulfilled if the magnitude of the inhomogeneities is not too high.

Let us conclude the analysis of the equilibrium distribution by outlining
two significant differences between the transport in statistically
homogeneous media and disordered photonic crystals. The most essential
difference is that the radiative transport in photonic crystals is provided
not by plane waves but by the modes, which reflect properties of the ideal
structure. Indeed, these are the modes of the photonic crystals that
constitute $\rho^{(\infty)}(\x_1,\x_2)$ rather than pure plane waves. As a
result, even in the asymptotic distribution of the energy density the
disorder does not wash out the features specific for the ideal structure. In
the case of weak disorder when the energy density can be approximated
by~(\ref{eq:energy_density_limit_PC}), one can see explicitly that
frequencies near the fundamental gap and higher, the spatial profile of
$w_{\kk, m}(\bi{R})$ is highly inhomogeneous on the length scale of the
elementary cell \cite{KUO:2004}. This inhomogeneity is stable, to some
extent, with respect to averaging over the equilibrium distribution
function. Most clearly this effect should be seen for frequencies near the
lower edge of the fundamental gap. Indeed, as follows from the variational
principle \cite{JOANNOPOULOS:1995}á at such frequencies the field
distribution takes higher values in the regions with higher values of the
refractive index for each point on the equifrequency surface. The same
obviously holds for the energy density of each mode. Thus, we can conclude
that even the long scale asymptotic behaviour of intensity in disordered
photonic crystals is highly inhomogeneous. This property is unexpected from
the point of view of the transport in statistically homogeneous media, where
the distinctive feature of this limit is the homogeneous distribution of the
energy density \cite{ISHIMARU:1978}þ

The second important feature of the field distribution in photonic
crystals is its significant anisotropy as a function of direction of the
Bloch vector. Anisotropy itself is not, of course, the specific feature of
disordered photonic crystals. The similar effect can be expected in
regular anisotropic media, where the group velocity depends on the
direction of the wave vector. In disordered photonic crystals, however,
this anisotropy can be very pronounced, especially near the frequencies
corresponding to the edges of (partial) band gaps.

\subsection{The radiative transfer equation}

Applying~(\ref{eq:averaged_Poynting}) to the equilibrium state described in
the previous subsection,
\begin{equation}\label{eq:flux_distribution_equilibrium}
  \bi{S}^{(\infty)}(\x_1,\x_2) =
  \widehat{\bi{S}}\rho^{(\infty)}(\x_1,\x_2),
\end{equation}
and using the symmetry properties of the equilibrium distribution given as
$\rho^{(\infty)}(\x_1,\x_2) = \rho^{(\infty)}(\x_2,\x_1)$, we can
immediately see that the flux in this state vanishes:
$\bi{S}^{(\infty)}(\bi{R},\bi{R}) = 0$. This is of course a result expected
for an asymptotic regime in the infinite medium. In order to describe states
with a non-zero flux, which are most relevant to experimental situations, we
have to consider the correlation function at a finite distance from the
sources or in a finite-sized medium.
In this case, we cannot expect that the same modes $\bar{\Psi}$ will
diagonalized the correlation function, but if the deviations from the
equilibrium state are small we can assume that non-diagonal elements decay
very fast when one moves away from the main diagonal of the density matrix.
Then, instead of trying to find the true diagonalizing states we will follow
the same approach as in the standard transport theory in statistically
homogeneous medium and convert the non-diagonal density-matrix into a slowly
changing in the real space function. The field-field correlator in this
representation takes the form of
\begin{equation}\label{eq:quasi_equilibrium_RTE}
  \rho(\x_1,\x_2) = \sum_{\kk,\bar{m}} \mathcal{I}_{\kk,\bar{m}}(\bi{R})
 \bar{\Psi}_{\kk,\bar{m}}(\x_1) \bar{\Psi}_{\kk,\bar{m}}^*(\x_2),
\end{equation}
where $\bi{R} = (\x_1 + \x_2)/2$ and $\mathcal{I}_{\kk,\bar{m}}(\bi{R})$ are
smooth functions of the
coordinate $\bi{R}$. 
The functions $\mathcal{I}_{\kk,\bar{m}}(\bi{R})$ can be interpreted as
specific intensities of the respective modes $\bar{\Psi}_{\kk,\bar{m}}$. In
order to show it let us calculate the average energy density and the average
Poynting vector. Using representation (\ref{eq:quasi_equilibrium_RTE}) in
Eqs.~(\ref{eq:average_energy_density}) and (\ref{eq:averaged_Poynting}) we
obtain
\begin{equation}\label{eq:average_energy_Poynting_RTE}
\eqalign{
 \aver{w(\bi{R})} =
 \sum_{\kk, \bar{m}}\mathcal{I}_{\kk,\bar{m}}(\bi{R})
 w_{\kk,\bar{m}}(\bi{R}), \cr
 \aver{\bi{S}(\bi{R})} =
 \sum_{\kk, \bar{m}}\mathcal{I}_{\kk,\bar{m}}(\bi{R})
 \bi{S}_{\kk,\bar{m}}(\bi{R}),}
\end{equation}
where we have used Eqs.~(\ref{eq:mu_mode_energy_density}) and
(\ref{eq:mu_mode_Poynting_vector}) to define formally the energy density,
$w_{\kk,\bar{m}}$ and the Poynting vector, $\bi{S}_{\kk,\bar{m}}$,
corresponding to the states $\bar{\Psi}_{\kk,\bar{m}}$.
Equations~(\ref{eq:average_energy_Poynting_RTE}) agree with the intuitive
concept of the specific intensity justifying our identification for
functions $\mathcal{I}_{\kk,\bar{m}}$.

In order to derive an equation for the specific intensities, we calculate
the matrix element of both sides of~(\ref{eq:rho_tansport_equation_steady})
between the states $\bar{\Psi}^*_{\kk+\qq/2,\bar{n}}$ and
$\bar{\Psi}_{\kk-\qq/2,\bar{n}}$ and then use the assumption of smooth
$\mathcal{I}_{\kk,\bar{m}}(\bi{R})$ as expressed
by~(\ref{eq:inverse_Fourier_smooth_density}). Calculations based on the
separation of length scales (see details in~\ref{sec:App_scales}) lead to
the radiative transfer equation in the form
\begin{equation}\label{eq:RTE_PC}
 \fl \omega\, {\bi{V}}_{\bar{m}}(\kk)\cdot \nabla_{\bi{R}}\mathcal{I}_{\kk,\bar{m}}(\bi{R})
  = \mathrm{Im}[\Sigma_{\bar{m},\bar{m}}(\kk)]\mathcal{I}_{\kk,\bar{m}}(\bi{R})
  + \int \rmd \qq\, \sum_{\bar{n}}\sigma_{\bar{m},\bar{n}}(\kk,
  \qq)\mathcal{I}_{\qq,\bar{n}}(\bi{R}).
\end{equation}
Here and below the matrix elements for the self-energy and the irreducible
vertex in the basis provided by the states $\bar{\Psi}_{\bar{\mu}}$ are
related to the matrix elements defined in
Eqs.~(\ref{eq:aveGreen_self-energy_coefficients}) and
(\ref{eq:irreducible_Bloch}) through transformation~(\ref{eq:PC_new_basis}).

In equation~(\ref{eq:RTE_PC}) we have introduced several important
quantities. The velocity ${\bi{V}}_{\bar{m}}(\kk)$ determines the direction
of the ``ray" corresponding to the mode $\bar{\Psi}_{\kk,\bar{m}}$ and is
expressed in terms of the group velocity of the photonic crystal modes as
\begin{equation}\label{eq:effective_group_velocity}
 \bi{V}_{\bar{m}}(\kk) = \frac{1}{\omega} \sum_n \left|B_{\bar{m},n}(\kk)\right|^2
 \omega_n(\kk) \bi{v}_n(\kk).
\end{equation}
The spatial decay of the initial distribution $\mathcal{I}_{\kk,m}(\bi{R})$
concentrated in a single mode towards equilibrium is quantified by the
mean-free-path
\begin{equation}\label{eq:mean_free_path_def}
  \ell_{\bar{m}}^{-1}(\kk) = -\frac{\mathrm{Im}[\Sigma_{\bar{m}, \bar{m}}(\kk)]}{\omega V_{\bar{m}}(\kk)}.
\end{equation}
As follows from the optical theorem, the decay is a result of the
redistribution of energy between the modes specified by different points
in the first Brillouin zone and different band numbers. Making use of
representation (\ref{eq:irreducible_Bloch}) the scattering kernel, which
describes the redistribution, can be shown to have the form
\begin{equation}\label{eq:scattering_kernel_RTE}
  \sigma_{\bar{m},\bar{n}}(\kk,\qq) = N \rho^{(\infty)}_{\bar{m}}(\kk)
  U^{\bar{m},\bar{m}}_{\bar{n},\bar{n}}(\kk, \kk, \qq, \qq).
\end{equation}

Initially, in relatively small samples or near the boundary of a big sample,
the distribution of energy can be arbitrary complex. For example, almost all
energy can be concentrated in only few modes. The distribution changes along
the sample according to~(\ref{eq:RTE_PC}) and eventually
$\mathcal{I}_{\kk,\bar{m}}(\bi{R})$ tends to $\rho^{(\infty)}_{\kk,
\bar{m}}$, which cancels both sides of~(\ref{eq:RTE_PC}). This confirms our
interpretation of $\rho^{(\infty)}_{\kk, \bar{m}}$ as the asymptotic
equilibrium distribution.

If we can neglect the interband coupling and use the on-shell approximation,
which is justifiable in the limit of weak and short-scale disorder, the
picture significantly simplifies since in this approximation
$B_{\bar{m},n}(\kk)$ become just identity matrices  and the density matrix
can be approximated by~(\ref{eq:density_equilibrium_group}). This means that
the transport is provided by the modes of the ideal photonic crystal
corresponding to the frequency $\omega$ and the specific intensity
$\mathcal{I}_{\kk, \bar{m}}$ has the transparent meaning of the intensity of
these modes. The spatial distribution of the specific intensity is governed
by the same equation~(\ref{eq:RTE_PC}) with the velocity
$\bi{V}_{\bar{m}}(\kk)$ substituted by the respective group velocity
$\bi{V}_{\bar{m}}(\kk) \to
\bi{v}_{m}(\kk)$ as follows from~(\ref{eq:effective_group_velocity}). 

\begin{figure}
  \includegraphics[width=5in]{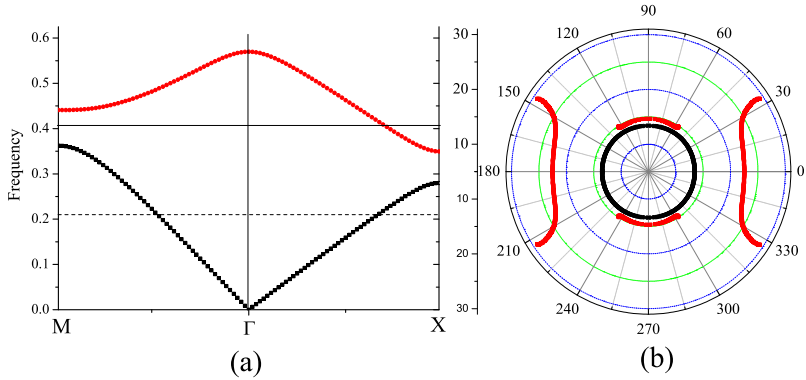}
  \caption{(Colour online) (a) The band structure of the ideal photonic crystal with the
  same parameters as in Figure~\ref{fig:group_velocity}. Only
  two bands used in calculations for Figs.~\ref{fig:group_velocity}, \ref{fig:scattering_kernel}b
  and \ref{fig:mean_free_path} are shown. The horizontal lines
  correspond to the frequencies $\omega a/2\pi = 0.21$ and $0.41$.
  (b) The scattering amplitude $\sigma(\kk, \qq)/\omega^3 V^2$ between the states belonging
  to the first ($\omega a/2\pi = 0.21$, the internal points) and the
  second ($\omega a/2\pi = 0.41$, the points corresponding to
  higher values) bands as a function of the angle between the vectors $\kk$ and $\qq$.
  The direction of the initial Bloch wave vector $\qq$ is taken along the $\Gamma X$
  direction for both frequencies, thus the angle in (b) is counted from the $\Gamma X$ direction.}\label{fig:scattering_kernel}
\end{figure}

In order to illustrate the specific features of the radiative transfer
equation in disordered photonic crystals, we calculate the scattering kernel
$\sigma_{\bar{m},\bar{n}}(\kk,\qq)$ for a 2D structure made of dielectric
discs in the limit of weak disorder. We approximate the intensity propagator
by the sum of the ladder diagrams and additionally assume the
$\delta$-correlated disorder when $K(\x_1, \x_2) = V^2\delta(\x_1 - \x_S)
\delta(\x_1-\x_2)$, where $\x_S$ denotes points lying in a thin shell near
the surface of the ideal disc [see~\ref{sec:App_random_model}]. In this case
the specific intensity is concentrated at the equifrequency surfaces
$F_m(\omega)$ so that it can be presented as $\mathcal{I}_{\kk, m}(\bi{R}) =
\widetilde{\mathcal{I}}_{\kk, m}(\bi{R})\delta\left[\omega^2 -
\omega_m^2(\kk)\right]$. The amplitudes $\widetilde{\mathcal{I}}_{\kk,
m}(\bi{R})$ satisfy the radiative transfer equation of the same structure
as~(\ref{eq:RTE_PC}) with the scattering amplitude between the states on the
equifrequency surface given by
\begin{equation}\label{eq:scattering_kernel_simplest}
  {\sigma}_{m,n}(\kk,\qq) = \frac{\omega^3V^2}{v_n(\qq)\mathcal{V}^2}
  \int_S \rmd\x\, |u_{\kk, m}(\x)|^2 |u_{\qq, n}(\x)|^2,
\end{equation}
where the integration is performed in a thin shell near the surface of the
ideal disc. We show the dependence of $\sigma_{m,n}(\kk,\qq)$ on the
direction of the final Bloch wave vector in
Figure~\ref{fig:scattering_kernel}b for two cases when the frequency
$\omega$ is well below the frequency of the fundamental gap $\omega_G$
($\omega_G\approx 0.28$ for the structure used in the numerical
calculations) and when $\omega \gtrsim \omega_G$. As one can see, for low
frequencies the scattering amplitude is close to isotropic (the variation of
the scattering amplitude is not visible in this scale), which is the
consequence of slow spatial variation of the Bloch amplitudes
$u_{\kk,m}(\x)$ and the shape of the equifrequency surface close to
spherical. However, when $\omega$ approaches $\omega_G$ both the spatial
variation of the Bloch amplitudes and the shape of the equifrequency surface
become nontrivial resulting in strongly anisotropic scattering. The
interesting feature of this dependence is that the scattering rates between
the states, which belong to the same band (the points corresponding to
$\omega a/2\pi = 0.41$ near the horizontal line), are noticeably higher than
the scattering amplitudes connecting the states in the different bands
(points near the vertical axis). It should be noted that since the gap is
not complete the scattering amplitude remains finite for all directions.

The expression for the mean-free-path $\ell_m(\kk)$ in terms of the
scattering amplitude $\sigma_{m,n}(\kk,\qq)$ is found from the requirement
for $\rho^{(\infty)}$ to solve the radiative transfer equation,
\begin{equation}\label{eq:inv_mean_free_path}
  \ell^{-1}_m(\kk) = \frac{1}{\omega v_m(\kk)}\sum_n \int \rmd\qq \, \sigma_{m,n}(\kk,\qq),
\end{equation}
where the integral runs over the respective equifrequency surfaces.
Substituting~(\ref{eq:scattering_kernel_simplest}) into this expression we
find $\ell^{-1}_m(\kk)$ in terms of the Bloch amplitudes. The same result
can be obtained from (\ref{eq:mean_free_path_def}) using for the self-energy
the same approximation as for
deriving~(\ref{eq:scattering_kernel_simplest}).

\begin{figure}
  \includegraphics[width=3in]{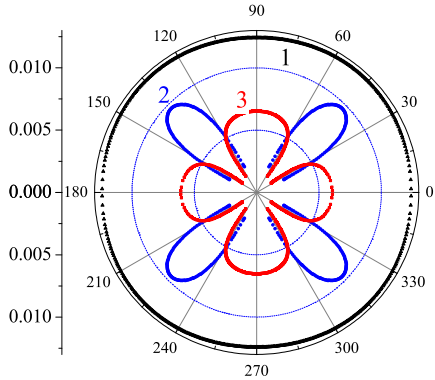}
  \caption{(Colour online) The directional dependence of the mean-free-path $\ell(\omega) \omega^4/V^2$
  for frequencies $\omega a/2\pi = 0.21$ (curve $1$),
 $0.31$ (curve $2$) and $0.41$ (curve $3$). The angle is counted from the $\Gamma X$-direction.
}\label{fig:mean_free_path}
\end{figure}

The evaluation of the mean-free-path for the same frequencies as in
Figure~\ref{fig:scattering_kernel} is presented in
Figure~\ref{fig:mean_free_path}. It is seen that for the relatively low
frequencies the mean-free-path only weakly depends on the direction of the
Bloch wave vector thus reproducing the well-known results for statistically
homogeneous media. In a vicinity of the partial band gap the anisotropic
scattering manifests in strong directional dependence of the mean-free-path.
This is especially evident for an intermediate frequency $\omega a/2\pi =
0.31$, for which near the $\Gamma M$-direction the mean-free-path drops by
the factor of $5$.

\subsection{The diffusion equation}

The right-hand side of~(\ref{eq:RTE_PC}) describing spatial variations of
the specific intensities consists of the sum of two terms. In order to
appreciate their physical significance let assume that at small distances
from the source only single mode is excited so that the specific intensity
can be presented in the following form
\begin{equation}\label{eq:I0}
\mathcal{I}_{\kk, m} \propto \mathcal{I}(R)\delta_{m,m_0}\delta(\kk-\kk_0).
\end{equation}
The first of the right hand terms in~(\ref{eq:RTE_PC}) describes exponential
spatial decay of this specific intensity due to scattering and
redistribution of the energy among other modes. At small distances this term
dominates and solution of~(\ref{eq:RTE_PC}) can be presented in the same
form as~(\ref{eq:I0}) with exponential spatial dependence. However, at
distances from the source exceeding the mean-free-path this term diminishes
and the second, integral term, starts determining the spatial distribution
of the specific intensities, which is no longer exponential. One can
describe this situation at asymptotically large distances taking into
account that in this limit the density matrix $\rho(\x_1,\x_2)$  must be
close to its limiting value $\rho^{(\infty)}(\x_1,\x_2)$ so that one can
present $\rho(\x_1,\x_2)$ using an appropriate expansion near
$\rho^{(\infty)}_{\kk,m}$. Drawing an analogy with the derivation of the
diffusion approximation in the standard case
\cite{LAGENDIJK:1996,ISHIMARU:1978,VAN_ROSSUM:1998} and assuming that the
photonic crystal has the centre of symmetry we present specific intensities
in the following form
\begin{equation}\label{eq:diff_approximation}
 \mathcal{I}_{\kk, m}(\bi{R}) = \frac{1}{v_m(\kk)}
 \left[W^{-1} w_{\mathrm{d}}(\bi{R}) +
 \widehat{\bi{v}}_m(\kk) \tensor{T}^{-1}\bi{S}_{\mathrm{d}}(\bi{R})\right],
\end{equation}
where $w_{\mathrm{d}}(\bi{R})$ and $\bi{S}_{\mathrm{d}}(\bi{R})$ are assumed
to be slowly changing functions of $\bi{R}$, while $W$ and $\tensor{T}$ are
scalar and tensor respectively, which do not depend on spatial coordinates.
With an appropriate choice of these quantities one can show that
$w_{\mathrm{d}}(\bi{R})$ and $\bi{S}_{\mathrm{d}}(\bi{R})$
in~(\ref{eq:diff_approximation}) can be presented as
\begin{equation}\label{eq:w_S_expressions}
 \eqalign{
 w_{\mathrm{d}}(\bi{R}) = \frac{1}{\mathcal{V}}\int_{\mathcal{V}(\bi{R})}
 \rmd\x\,w(\x,\x), \cr
 \bi{S}_{\mathrm{d}}(\bi{R}) = \frac{1}{\mathcal{V}}\int_{\mathcal{V}(\bi{R})}
 \rmd\x\, \bi{S}(\x, \x),}
\end{equation}
where the integration is performed over the elementary cell with the
coordinate $\bi{R}$; $w(\x_1, \x_2)$, and $\bi{S}(\x_1, \x_2)$ are defined
by Eqs.~(\ref{eq:energy_generating}) and (\ref{eq:flux_distribution}),
respectively. These expressions clarify the physical meaning of
$w_d(\bi{R})$ and $\bi{S}_d(\bi{R})$ as long scale envelopes of the averaged
energy density and the flux, respectively. $W$ and $\tensor{T}$
in~(\ref{eq:diff_approximation}) are found to be
\begin{equation}\label{eq:constants_diffusion}
  W = \frac{\omega}{2} \lambda(\omega), \qquad
 \tensor{T} = \frac{\omega}{2} \sum_m \int \rmd\kk \,
 \widehat{\bi{v}}_m(\kk) \otimes \widehat{\bi{v}}_m(\kk).
\end{equation}
where $\otimes$ denotes the tensor product, $(\bi{v}\otimes \bi{u})_{ij} =
v_i u_j$ and $\lambda(\omega)$ is the density of states of the photonic
crystal defined as  $\lambda(\omega) = \sum_\mu \delta(\omega -
\omega_\mu)$.

The equations with respect to $w_{\mathrm{d}}(\bi{R})$ and
$\bi{S}_{\mathrm{d}}(\bi{R})$ can be derived from the radiative transfer
equation. Summation over all states on the respective equifrequency surface
yields the energy conservation law
\begin{equation}\label{eq:energy_cons_diff}
  \nabla \cdot \bi{S}_{\mathrm{d}}(\bi{R}) = 0.
\end{equation}
Multiplying both sides of~(\ref{eq:RTE_PC}) by $\widehat{\bi{v}}_m(\kk)$ and
summing over all states we obtain
\begin{equation}\label{eq:Flicks_law_diff}
  \nabla\tensor{D}\nabla w_{\mathrm{d}}(\bi{R}) = 0,
\end{equation}
where the diffusion tensor $\tensor{D}(\omega)$ up to a constant factor is
defined  by $\tensor{D}(\omega) = \tensor{T}\tensor{M}^{-1}\tensor{T}$ with
the current relaxation kernel \cite{VOLLHARDT:1980}
\begin{equation}\label{eq:M_definition}
\fl  \tensor{M} = \frac{\omega^2 \lambda(\omega)}{4} \sum_{m}
  \int d\kk \, \left[ \ell^{-1}_m(\kk) -
  \sum_n \int \rmd\qq\,\frac{\sigma_{mn}(\kk,\qq)}{\omega v_m(\kk)}
  \widehat{\bi{v}}_m(\kk) \otimes \widehat{\bi{v}}_n(\qq)
  \right].
\end{equation}

It is important to note that despite the scalar character of the initial
Helmholtz equation~(\ref{eq:Helmholtz_equation}) the diffusion in photonic
crystals is characterized by a diffusion \textit{tensor} rather than by a
single diffusion constant. In particular, if the periodic modulation is
characterized by a rectangular (noncubic) elementary cell then the
eigenvalues of the tensor corresponding to different principal directions
will be different resulting in anisotropic diffusion.

\begin{figure}
  \includegraphics[width=3in]{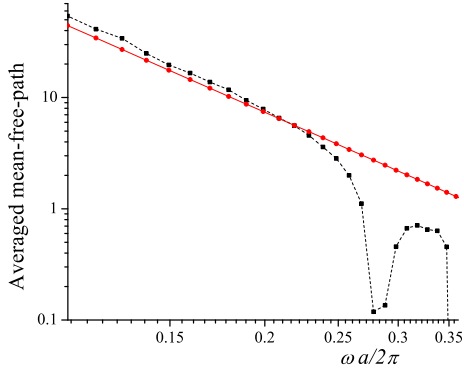}
  \caption{(Colour online) The dependence of the effective mean-free-path $\ell(\omega)/V^2$ (dashed line, squares)
  on frequency in log-log scale. The solid line (circles) shows
  the dependence $\propto \omega^3$ (this asymptotics for
  low frequencies is due to the dimensionality of the problem).}\label{fig:sym_mfp}
\end{figure}

This diffusion anisotropy, however, appears only in low symmetry photonic
crystals. If the point symmetry group of a particular structure is such that
irreducible representations of the full rotation group remain irreducible
for the point group of the crystal, the diffusion tensor is reduced to a
scalar form. Indeed, crystals with such symmetries, e.g. crystals with
square and hexagonal lattices in 2D and cubic and fcc lattices in 3D, do not
allow non-trivial tensors of the second rank \cite{Bir_Pikus_Symmetries}þ In
this case the diffusion, despite of the strong directional dependence of the
scattering kernel $\sigma_{mn}(\kk,\qq)$, is isotropic and is characterized
by the diffusion constant
\begin{equation}\label{eq:diff_constant_sym}
  D = \frac{F(\omega)}{d \lambda(\omega)} \,\frac{\ell(\omega)}{1-\aver{\cos(\theta)}},
\end{equation}
where $d$ is the dimensionality of the problem, $F(\omega) = \sum_m \int
d\kk\, 1$ is the total area of the equifrequency surface,
\begin{equation}\label{eq:single_ell_def}
  \ell^{-1}(\omega) = \frac 1{F(\omega)}\sum_m \int \rmd\kk\, \ell_m^{-1}(\kk)
\end{equation}
and
\begin{equation}\label{eq:aver_cos}
  \aver{\cos(\theta)} \equiv \frac{\ell(\omega)}{F(\omega)} \sum_{m,n}
  \int \rmd\kk \rmd\qq\, \frac {\sigma_{mn}(\kk,\qq)}{\omega v_m(\kk)} \,
  \widehat{\bi{v}}_{m}(\kk)\cdot\widehat{\bi{v}}_{n}(\qq).
\end{equation}
The expression for the diffusion coefficient has the same general structure
as for statistically homogeneous media \cite{LAGENDIJK:1996} with the
transport velocity $v_E(\omega) = F(\omega)/\lambda(\omega)$ and the
``transport mean-free-path" $\ell(\omega)/(1-\aver{\cos \theta})$. The
transport velocity is independent on the details of the disorder
distribution owing to the $\delta$-functional approximation for the
correlation function $K(\x_1, \x_2)$, which is valid if the frequencies
corresponding to morphological resonances at the typical inhomogeneity size
is much higher than the range of frequencies under consideration
\cite{VAN:1992:ID3086,BARABANENKOV:1992:ID3303,VAN:1993:ID3301}.
Figure~\ref{fig:sym_mfp} shows the frequency dependence of $\ell(\omega)$
for a disordered photonic crystal with the square lattice. It should be
noted that the minimum of $\ell(\omega)$ is reached at the band edge in the
$\Gamma X$-direction (see Figure~\ref{fig:scattering_kernel}a) and is due to
the significant drop of the group velocity when the equifrequency surface
touches the boundary of the first Brillouin zone.

It should be emphasized, however, that the possibility to introduce a single
scalar diffusion constant and, respectively, to describe transport by a
single transport mean-free-path is not a a result of the disorder destroying
the effect of periodicity but rather is the consequence of the underlying
symmetry of the photonic crystal. In the case of low symmetry structures the
diffusion is described by a tensor, and is characterized by different
effective mean-free-paths for different principle directions of the
diffusion tensor.

\section{Conclusion}
\label{sec:discussion}

In this paper we developed a systematic approach based on multiple
scattering analysis to the theoretical description of incoherent transport
properties of disordered photonic crystals in the steady state regime. The
main difficulty in developing such a theory using a standard plane-wave
based multiple scattering approach lies in the separation between waves
scattering due to periodic modulations of the dielectric constant and
scattering due to random deviations from the periodicity. One might think
that this difficulty is readily resolved by incorporating modes of the
ideal photonic crystal instead of plane waves into the standard transport
theory. However, such a straightforward approach fails because of the
inherent presence of two spatial scales, the period of the crystal and the
main-free-path, both of which has to be retained in the theory. As a
result, a formally constructed Wigner distribution, which is the main
technical tool in deriving the radiative transfer equation in the regular
case of statistically homogeneous medium, loses its smooth spatial
dependence and, hence, its methodological usefulness.

We showed that these difficulties can be overcome if one incorporates the
concept of the radiative transfer as a incoherent process into the
foundation of the theory. In order to achieve this, we showed that the
field-field correlation function, $\rho(\x_1, \x_2) =
\aver{E(\x_1)E^*(\x_2)}$, has all the properties of a density matrix
defined in quantum statistics, and hence, can be interpreted as such. This
interpretation allowed us to separate coherent and incoherent
contributions to the transport and eventually obtain radiative transfer
and diffusion equations for media with periodically modulated average
dielectric function.

We found an exact asymptotic solution, $\rho^{(\infty)}$, of the
Bethe-Salpeter equation~(\ref{eq:density_limit}), which describes the
limiting form of the correlation function in an infinite medium
asymptotically far away from the sources. This function determines the
asymptotic distribution of the intensity of the field, which was shown to be
highly spatially non-uniform and anisotropic. This result should be
contrasted with the case of statistically homogeneous medium, in which the
asymptotic distribution of intensity is spatially uniform. It is important
that the properties of the intensity distribution even inside an infinite
photonic crystal are determined by the normal modes of the underlying
periodic structure. This result casts serious doubts on the assumption made
in Refs.~\cite{KOENDERINCK:2003,KOENDERINCK:2005} that the field
distributions deep inside a photonic crystal and a statistically homogeneous
media do not differ from each other, and that the anisotropy of the light
emerging from photonic crystals is formed within a narrow layer at the
boundary.

The asymptotic character of $\rho^{(\infty)}$ is expressed by the absence of
flux in this state. Using analogy with the statistical physics we can
interpret $\rho^{(\infty)}$ as an equilibrium distribution. In order to
describe actual energy transfer, one needs to consider small deviations from
equilibrium. We derived generalized forms of radiative transfer and
diffusion equations valid for disordered photonic crystals and obtained
general expressions for the scattering cross-section, mean free path, and
the diffusion coefficient. As an example, we calculated the cross-section
for one particular model of a photonic crystal and demonstrated how the
underlying periodic structure effects disorder induced scattering of
photonic modes.  In particular, we found that the scattering cross-section
describing the redistribution of the energy between modes becomes highly
anisotropic at high frequencies. The mean-free-path of a particular mode,
which describes the rate of redistributing the correspondent specific
intensity, also depends strongly on the direction of the respective Bloch
vector. The numerical calculations show that for a chosen frequency near the
(partial) band gap the mean-free-path may vary almost by the order of
magnitude. Such significant variation presents especial interest from the
perspective of the problem of the Anderson localization.

The interesting feature of the transport in the disordered photonic crystals
is that even if scattering between different modes may be highly anisotropic
it does not necessarily imply an anisotropic \textit{diffusion}. Indeed, if
the point symmetry of the crystal is sufficiently rich, e.g. in 2D this is
rotational symmetry of the third order or higher, then all tensors of the
second rank describing the intrinsic macroscopic properties are proportional
to the unit tensor. That is the diffusion is necessarily isotropic and is
characterized by a single diffusion constant. The mean-free-path appears in
this constant averaged over the respective equifrequency surface.

The derivation of the transport equations can be generalized to describe the
radiative transport in more general situations than considered in the paper.
Principally, the equilibrium distribution given by $\rho^{(\infty)}$ holds
in the presence of the boundary as well. The latter is accounted through the
boundary conditions determining Green's functions. However, whether the
respective density matrix can be described near the boundary as a long-scale
perturbation of $\rho^{(\infty)}$ or not presents a special interesting
problem.

Finally, we would like to comment on the possibility to extend the
consideration provided in the paper to a non-steady-state regime. By a
direct analogy with the standard case, a slow dynamics in the diffusion
regime can be accounted for by assuming $w_d$
in~(\ref{eq:diff_approximation}) to be time dependent. However, a rigourous
theory, which would yield this limit is yet to be developed. From the
perspectives of the presented consideration, the main formal obstacle is the
fact that $\rho_{\omega_1, \omega_2}$ is generally not non-negatively
defined if $\omega_1 \ne \omega_2$. This problem, of course, exists and is
recognized in the standard theory of multiple scattering in homogeneous
media as well. In order to ensure positivity of the specific intensities, it
was suggested to consider a coarse-grained Wigner distribution
\cite{JOHN:1996}. A possibility to generalize our approach to time-dependent
correlation functions and their possible relation with the density matrix
formalism is an open problem.

\ack

The work of the Northwestern University group is supported by
National Science Foundation under the Grant No.~DMR-0093949, and the
work of Queens College group is supported by AFOSR under the Grant
No.~FA9550-07-1-0391 and PCS-CUNY grants.

\appendix

\section{A model of Gaussian inhomogeneities in photonic crystals}
\label{sec:App_random_model}

A Gaussian random field $\Delta\epsilon(\x)$ can be defined in two
equivalent ways. The functional definition is based on the requirement of an
integral
\begin{equation}\label{eq:functional_Gaussian}
  I_f = \int \rmd\x\, f(\x) \Delta\epsilon(\x)
\end{equation}
to be a Gaussian random variable for an arbitrary function $f(\x)$.
The statistical definition states that a multi-point correlation
function of a zero-mean Gaussian field is expressed in terms of the
two-point correlation function
\begin{equation}\label{eq:gauss_general_rule}
  \aver{\Delta\epsilon(\x_1)\ldots\Delta\epsilon(\x_{2n})} =
  \sum_{\mathrm{pairs}}K(\x_\alpha,\x_\beta)\ldots
  K(\x_\gamma,\x_\delta),
\end{equation}
where the summation is taken over all possible pairings of the points
$\x_1,\ldots, \x_{2n}$ and $K(\x_1, \x_2) =
\aver{\Delta\epsilon(\x_1)\Delta\epsilon(\x_2)}$.

Property of integral~(\ref{eq:functional_Gaussian}) being a Gaussian random
variable is consistent with the intuitive idea about a Gaussian random
field, while property~(\ref{eq:gauss_general_rule}) is convenient from the
technical point of view for developing the perturbation theory
\cite{Rytov_Principles_IV}. Both these properties, however, might seem
somewhat artificial from the perspective of disordered photonic crystal.
Here we would like to present a simple model of the disorder in a photonic
crystal that naturally results in simulation of the inhomogeneities by a
Gaussian random field.

Let the ideal structure be constituted by spheres with the dielectric
constant $\epsilon_1$ at the sites of a periodic lattice. Furthermore, we
assume that the disorder in this system is constituted by slight
variations of the size of the spheres, while their positions are fixed.
Thus, we can represent the spatial modulation of the dielectric function
in the disordered crystal in the form
\begin{equation}\label{eq:index_spheres_disordered}
  \widetilde{\epsilon}(\x) = \epsilon_0 +
  \sum_{\bi{R}}\delta \epsilon_{\bi{R}}(\x + \bi{R}),
\end{equation}
where $\bi{R}$ are the lattice vectors, $\epsilon_0$ is the background
dielectric constant, and $\delta \epsilon_{\bi{R}}(\x + \bi{R})$ is the
deviation from the background value in the elementary cell with the
coordinate $\bi{R}$. By the construction $\delta \epsilon_{\bi{R}}(\x) =
\epsilon_1 - \epsilon_0$ inside the sphere and it equals to $0$ elsewhere.
We represent the spatial distribution of the dielectric function in the form
adapted in~(\ref{eq:Helmholtz_equation}), i.e. $\widetilde{\epsilon}(\x) =
\epsilon(\x) + \Delta\epsilon(\x)$, where $\epsilon(\x)$ describes the
distribution in the ideal structure and
\begin{equation}\label{eq:random_add_spheres}
  \Delta\epsilon(\x) = \sum_{\bi{R}}\left(\delta \epsilon_{\bi{R}}(\x +
  \bi{R}) - \delta \epsilon_0(\x + \bi{R})\right)
\end{equation}
with $\delta \epsilon_0(\x)$ being the deviation from the background value
in the ideal structure.

Clearly $\Delta\epsilon(\x)$ is not zero in shells near the boundaries of
the spheres constituting the ideal structure. It is positive in the spheres
with the radius larger than the radius of the ideal spheres and is negative
in smaller spheres. We simulate this function by presenting it in the form
\begin{equation}\label{eq:random_model_assumption}
  \Delta\epsilon(\x) = \sum_{\bi{R}}\Delta\epsilon_{\bi{R}} u(\x+\bi{R}),
\end{equation}
where $u(\x)$ is a non-random function different from zero in a thin shell
near the boundary of the ideal sphere and the random amplitudes
$\Delta\epsilon_{\bi{R}}$ have the Gaussian distribution with zero mean
value and are independent in different elementary cells.
Using~(\ref{eq:random_model_assumption}) in the integral
in~(\ref{eq:functional_Gaussian}) one has a Gaussian random variable and,
hence, $\Delta\epsilon(\x)$ is the Gaussian random field.

This model, obviously, can be generalized by allowing the random variables
$\Delta\epsilon_{\bi{R}}$ to be a (usual) Gaussian random field rather than
a constant inside each elementary cell. Such model of the disorder would
account not only for the size dispersion of the spheres but also for the
roughness of their surfaces. Assuming that $\Delta\epsilon_\x$ in different
elementary cells are not correlated we obtain
\begin{equation}\label{eq:generalized_correlation}
  K(\x_1, \x_2) = \aver{\Delta\epsilon_{\x_1}\Delta\epsilon_{\x_2}} u(\x_1)u(\x_2)
\end{equation}
if $\x_1$ and $\x_2$ are situated inside the same elementary cell and
$K(\x_1, \x_2) = 0$ otherwise. For calculations in the main text we have
used the simplest model of spheres with uniform sizes and with centers at
the sites of an ideal lattice but with rough surfaces. Thus we take $u(\x) =
1$ in a thin shell around the ideal sphere (respectively, the circle in 2D)
and
\begin{equation}\label{eq:sub_correlation}
  \aver{\Delta\epsilon_{\x_1}\Delta\epsilon_{\x_2}} = V^2\delta(\x_1 - \x_2).
\end{equation}

\section{Symmetries of the averaged Green's function, the self-energy and the irreducible vertex}
\label{sec:App_symmetries}

Leaving the problem of convergence of the perturbational series aside, the
symmetry properties of the averaged Green's function $\aver{G(\x_1, \x_2)}$,
the self-energy $\Sigma(\x_1, \x_2)$, and the irreducible vertex $U(\x_1,
\x_2; \x'_1, \x'_2)$ can be proven on the diagram by diagram basis
\cite{Rytov_Principles_IV}. We demonstrate the typical line of arguments by
proving that if the correlation function of inhomogeneities is invariant
with respect to lattice translations then so is the averaged Green's
function.

Any diagram in the perturbational expansion of $\aver{G(\x_1, \x_2)}$ has
the form of a line with $2n$ internal ($n\geq 0$) and $2$ terminating
vertices. The internal vertices divide the line on $2n+1$ segments, which
are the graphical representations of the non-perturbed Green's function of
the ideal structure. The internal vertices are connected pair-wise by the
lines of the correlation function.

Clearly the value of a diagram with fixed values of the coordinates
corresponding to the vertices does not change if we shift the coordinates of
all points (internal and terminating) by the vector of the lattice
translation. Indeed, the Green's function lines and the lines corresponding
to the correlation functions do not change because of the translational
invariance of these functions. Furthermore, we note that in order to obtain
the contribution of the diagram into $\aver{G(\x_1, \x_2)}$ we need to
integrate with respect to coordinates of the internal vertices. Finally, we
observe that the uniform shift of the coordinates of all vertices produces
the diagram, which corresponds to the perturbational series for
$\aver{G(\x_1 +\bi{R}, \x_2 + \bi{R})}$.

This line of arguments proves the translational invariance of the averaged
Green's function, the self-energy and the irreducible vertex. In order to
prove the reciprocity of the Green's function and the self-energy
\begin{equation}\label{eq:reciprocities_GF_SE}
  \aver{G(\x_1, \x_2)} = \aver{G(\x_2, \x_1)}, \qquad
  \Sigma(\x_1, \x_2) = \Sigma(\x_2, \x_1),
\end{equation}
we need the version of the arguments sketched above, which includes the
directionality of the lines constituting the diagram. The reciprocity
would follow then from invariance of the diagram with respect to the reversion
of the directionality of all lines. For the irreducible vertex we
additionally need to take into account that the correlation function
connecting upper and lower lines is mapped to itself upon the reverting.
Thus, we have
\begin{equation}\label{eq:IV_reciprocity}
  U(\x_1, \x_2; \x'_1, \x'_2) = U(\x'_1, \x'_2; \x_1, \x_2).
\end{equation}

We complete the consideration of the symmetry properties by proving the
invariance of the self-energy with respect to the point symmetries of the
photonic crystal. More specifically, we show that if the correlation
function transforms according to the identity representation of the
respective group, then so does the self-energy. The proof is based on the
property of the unperturbed Green's function $G_0(\x_1, \x_2)$ to
transform according to the identity representation
\begin{equation}\label{eq:identity_GF}
  \mathtt{T}\, G_0(\x_1, \x_2) \equiv G_0(\mathtt{T}^{-1}\,\x_1,
  \mathtt{T}^{-1}\,\x_2) = G_0(\x_1, \x_2),
\end{equation}
where $\mathtt{T}$ is an element of the point symmetry group. The proof of
the invariance of the self-energy uses the same arguments as above with
the only difference that now instead of shift or inversion operators we
act with $\mathtt{T}$ on all points of the diagram.

It follows from this consideration that the eigenfunctions of the Dyson
equation can be classified according to the irreducible representations of
the group of the point symmetries of the ideal structure. We use this fact
in the paper when we discuss the effect of the disorder on the
degenerate points in the spectrum of the photonic crystal.

\section{Separation of spatial scales in close-to-equilibrium regimes}
\label{sec:App_scales}

The main assumption used for the derivation of the radiative transfer
equation and the diffusion equation in the paper is that the spatial
scales of variation of the specific intensity or the envelope of the
energy density and of the functions describing the modes of the ideal
photonic crystal can be separated. Physically this assumption implies that
the slowly varying functions do not lead to coupling between different
Bloch modes. As a result, one can neglect the smooth functions while
calculating the respective scalar products.

In order to give a formal expression for the separation of length scales we
consider a photonic crystal mode $\Psi_{\kk,m}(\x)$ modulated by a smooth
function $f(\x)$
\begin{equation}\label{eq:Bloch_modulated}
  h_{\kk,m}(\x) = f(\x)\Psi_{\kk,m}(\x).
\end{equation}
The smoothness of the function $f(\x)$ can be quantified by
$\aver{\bi{p}}_f$ and $\aver{p^2}_f$, where
\begin{equation}\label{eq:smoothness_parameters}
  \aver{\bi{p}}_f = \int \rmd\bi{p}\, \bi{p}f({\bi{p}})
\end{equation}
and $f({\bi{p}})$ is the Fourier image of $f(\x)$. The limit of smooth
$f(\x)$ can, thereby, be formalized as $\aver{\bi{p}}_f \to 0$ and
$\aver{p^2}_f \to 0$. In this limit the weighted scalar product of
$h_{\kk,m}(\x)$ with the mode $\Psi_{\qq,n}(\x)$ is written as
\begin{equation}\label{eq:weighted_product}
 \left(\Psi_{\qq,n}, h_{\kk,m}\right)
 \equiv\int \rmd\x\, \epsilon(\x)\Psi^*_{\qq,n}(\x) h_{\kk,m}(\x) =
 f({\kk-\qq})\mathcal{U}_{n,m}(\qq, \kk),
\end{equation}
where
\begin{equation}\label{eq:U_matrix}
  \mathcal{U}_{n,m}(\qq, \kk) = \int_{\mathcal{V}} \rmd\x\,
  \epsilon(\x)u^*_{\qq,n}(\x)u_{\kk,m}(\x).
\end{equation}
The inverse Fourier transform of~(\ref{eq:weighted_product}) with respect to
$\kk-\qq$ gives for $m=n$
\begin{equation}\label{eq:inverse_Fouries}
  \int \rmd\bi{p}\, \left(\Psi_{\kk-\bi{p}/2,m}, h_{\kk +
  \bi{p}/2,m}\right) \rme^{-\rmi \bi{p}\cdot \x} = f(\x) +
  O(f'(\x)),
\end{equation}
where the remainder is small together with the derivatives of
$f(\x)$.

Similarly, the notion of the smooth modulation can be applied for
functions of the form
\begin{equation}\label{eq:smooth_density_modulation}
  \rho(\x_1, \x_2) = \sum_{\kk, m} \mathcal{I}_{\kk,m}\left[(\x_1 + \x_2)/2\right]
  \Psi_{\kk,m}(\x_1)\Psi^*_{\kk,m}(\x_2),
\end{equation}
where $\mathcal{I}_{\kk,m}(\x)$ are smooth functions. The weighted matrix
element of $\rho(\x_1, \x_2)$ between $\Psi_{\bi{p}+\qq/2,n}$ and
$\Psi_{\bi{p}-\qq/2,n}$ is
\begin{equation}\label{eq:rho_matrix_element}
\fl \left(\Psi_{\bi{p}+\qq/2,n} |\rho
 |\Psi_{\bi{p}-\qq/2,n}\right)  =
  \sum_m \mathcal{I}_{\bi{p}, m}(\bi{q}) \mathcal{U}_{m,n}(\bi{p}+\qq/2, \bi{p})
  \mathcal{U}_{n,m}(\bi{p},\bi{p}-\qq/2).
\end{equation}
The inverse Fourier transform with respect to $\qq$ gives
\begin{equation}\label{eq:inverse_Fourier_smooth_density}
  \int \rmd\qq\, \rme^{-\rmi \qq \cdot
  \bi{R}}\left(\Psi_{\bi{p}+\qq/2,n}\right|\rho\left|\Psi_{\bi{p}-\qq/2,n}\right)
  = \mathcal{I}_{\bi{p},m}(\bi{R})
\end{equation}
up to terms vanishing with the derivatives of
$\mathcal{I}_{\kk,m}(\x)$.

Additionally, for derivation of the radiative transfer equation one needs to
compare the scales related to the irreducible vertex. Analyzing the
diagrammatic expansion for $U(\x_1, \x_2; \x_1', \x_2')$ one can see that
there are two typical spatial scales determining the decay of the
irreducible vertex with the distance between the first and the second pairs
of the points. The ``short" scale is proportional to the correlation radius
of the inhomogeneities. This spatial decay is characteristic for ladder and
maximal crossed diagrams approximations. Assuming that the correlations
vanish at the scale of the elementary cell one can approximate
\begin{equation}\label{eq:scale_separation_I}
\eqalign{\fl \int \rmd\x_1' \rmd\x_2' \, U_{\omega_1,\omega_2} & (\x_1, \x_2; \x_1',
  \x_2') \mathcal{I}\left[(\x_1' + \x_2')/2\right]  \ldots \cr
  & \approx \mathcal{I}\left[(\x_1 + \x_2)/2\right]
  \int \rmd\x_1' \rmd\x_2' \, U_{\omega_1,\omega_2}(\x_1, \x_2; \x_1',
  \x_2') \ldots .}
\end{equation}
This approximation has been used for derivation of~(\ref{eq:RTE_PC}).

There are additional terms in the diagrammatic expansion of the
irreducible vertex that go beyond the ladder and maximally crossed
diagrams approximations and lead to the spatial decay on the scale
of the mean-free path. These terms would result in a non-local term
in the radiative transfer equation. The effect of the non-local
scattering on the radiative transfer in disordered photonic crystals
will be studied elsewhere. 

\section*{References}


\begin{thebibliography}{10}

\bibitem{JOANNOPOULOS:1995}
J.~D. Joannopoulos, R.~D. Meade, and J.~N. Winn.
\newblock {\em Photonic crystals: molding the flow of light}.
\newblock Princeton Univ. Press, Princeton, 1995.

\bibitem{HO:1990}
K.~M. Ho, C.~T. Chan, and C.~M. Soukoulis.
\newblock Existence of a photonic gap in periodic dielectric structures.
\newblock {\em Phys. Rev. Lett.}, 65(25):3152--3155, 1990.

\bibitem{KOSAKA:1998}
H.~Kosaka, T.~Kawashima, A.~Tomita, M.~Notomi, T.~Tamamura, T.~Sato, and
  S.~Kawakami.
\newblock Superprism phenomena in photonic crystals.
\newblock {\em Phys. Rev. B}, 58(16):R10096--R10099, 1998.

\bibitem{NOTOMI:2000}
M.~Notomi.
\newblock Theory of light propagation in strongly modulated photonic crystals:
  refractionlike behavior in the vicinity of the photonic band gap.
\newblock {\em Phys. Rev. B}, 62(16):10696--10705, 2000.

\bibitem{GUVEN:2004}
K.~Guven, K.~Aydin, K.~B. Alici, C.~M. Soukoulis, and E.~Ozbay.
\newblock Spectral negative refraction and focusing analysis of a
  two-dimensional left-handed photonic crystal lens.
\newblock {\em Phys. Rev. B}, 70:205125--5, 2004.

\bibitem{ILIEW:2005}
R.~Iliew, C.~Etrich, and F.~Lederer.
\newblock Self-collimation of light in three-dimensional photonic crystals.
\newblock {\em Opt. Express}, 13(18):7076--7085, 2005.

\bibitem{SHIN:2005}
J.~Shin and S.~Fan.
\newblock Conditions for self-collimation in three-dimensional photonic
  crystals.
\newblock {\em Opt. Lett.}, 30(18):2397--2399, 2005.

\bibitem{PELEG:2007}
O.~Peleg, G.~Bartal, B.~Freedman, O.~Manela, M.~Segev, and D.~N.
  Christodoulides.
\newblock Conical diffraction and gap solitons in honeycomb photonic lattices.
\newblock {\em Phys. Rev. Lett.}, 98:103901--103901--4, 2007.

\bibitem{ROBERTSON:1992}
W.~M. Robertson, G.~Arjavalingam, R.~D. Meade, K.~D. Brommer, A.~M. Rappe, and
  J.~D. Joannopoulos.
\newblock Measurement of photonic band structure in a two-dimensional
  dielectric array.
\newblock {\em Phys. Rev. Lett.}, 68(13):2023--2026, 1992.

\bibitem{SAKODA:1995}
K.~Sakoda.
\newblock Symmetry, degeneracy, and uncoupled modes in two-dimensional photonic
  lattices.
\newblock {\em Phys. Rev. B}, 52(11):7982--7986, 1995.

\bibitem{KARATHANOS:1998}
V.~Karathanos.
\newblock Inactive frequency bands in photonic crystals.
\newblock {\em J. Mod. Opt.}, 45(8):1751--1758, 1998.

\bibitem{SIGALAS:1999}
M.~M. Sigalas, C.~M. Soukoulis, C.~T. Chan, R.~Biswas, and K.~M. Ho.
\newblock Effect of disorder on photonic band gaps.
\newblock {\em Phys. Rev. B}, 59(20):12767--12770, 1999.

\bibitem{LI:2000}
Z.-Y. Li and Z.-Q. Zhang.
\newblock Fragility of photonic band gaps in inverse-opal photonic crystals.
\newblock {\em Phys. Rev. B}, 62(3):1516--1519, 2000.

\bibitem{KALITEEVSKI:2006}
M.~A. Kaliteevski, D.~M. Beggs, S.~Brand, R.~A. Abram, and V.~V. Nikolaev.
\newblock Stability of the photoic band gap in the presence of disorder.
\newblock {\em Phys. Rev. B}, 73:033106--033106--4, 2006.

\bibitem{KOENDERINCK:2003}
A.~F. Koenderink and W.~L. Vos.
\newblock Light exiting from real photonic band gap crystals is diffusive and
  strongly directional.
\newblock {\em Phys. Rev. Lett.}, 91(21):213902--213902--4, 2003.

\bibitem{KOENDERINCK:2005}
A.~F. Koenderink and W.~L. Vos.
\newblock Optical properties of real photonic crystals: anomalous diffuse
  transmission.
\newblock {\em J. Opt. Soc. Am. B}, 22(5):1075--1084, 2005.

\bibitem{KOENDERINCK:2000}
A.~F. Koenderink, M.~Megens, G.~van Soest, W.~L. Vos, and A.~Lagendijk.
\newblock Enhanced backscattering from photonic crystals.
\newblock {\em Phys. Lett. A}, 268:104--111, 2000.

\bibitem{HUANG:2001}
J.~Huang, N.~Eradat, M.~E. Raikh, Z.~V. Vardeny, A.~A. Zakhidov, and R.~H.
  Baughman.
\newblock Anomalous coherent backscattering of light from opal photonic
  crystals.
\newblock {\em Phys. Rev. Lett.}, 86(21):4815--4818, 2001.

\bibitem{SIVACHENKO:2001}
A.~Y. Sivachenko, M.~E. Raikh, and Z.~V. Vardeny.
\newblock Coherent umklapp scattering of light from disordered photonic
  crystals.
\newblock {\em Phys. Rev. B}, 63:245103--245103--7, 2001.

\bibitem{YAMILOV:2003}
A.~Yamilov and H.~Cao.
\newblock Density of resonant states and a manifestation of photonic band
  structures in small clusters of spherical particles.
\newblock {\em Phys. Rev. B}, 68:085111--085111--5, 2003.

\bibitem{KOENDERINCK:2005_mfp}
A.~F. Koenderink, A.~Lagendijk, and W.~L. Vos.
\newblock Optical extinction due to intrinsic structural variations of photonic
  crystals.
\newblock {\em Phys. Rev. B}, 72:153102--153102--4, 2005.

\bibitem{RYZHIK:1996}
L.~Ryzhik, G.~Papanicolaou, and J.~B. Keller.
\newblock Transport equations for elastic and other waves in random media.
\newblock {\em Wave Motion}, 24:327--370, 1996.

\bibitem{FERWERDA:1999}
H.~A. Ferwerda.
\newblock The radiative transfer equation for scattering media with a spatially
  varying refractive index.
\newblock {\em J. Opt. A: Pure Appl. Opt.}, 1:L1--L2, 1999.

\bibitem{SHENDELEVA:2004}
M.~L. Shendeleva.
\newblock Radiative transfer in a turbid medium with a varying refractive
  index: comment.
\newblock {\em J. Opt. Soc. Am. A}, 21(12):2464--2467, 2004.

\bibitem{BAL:2006}
G.~Bal.
\newblock Radiative transfer equations with varying refractive index: a
  mathematical perspective.
\newblock {\em J. Opt. Soc. Am. A}, 23(7):1639--1644, 2006.

\bibitem{Rytov_Principles_IV}
S.~M. Rytov, Yu.~A. Kravtsov, and V.~I. Tatarskii.
\newblock {\em Wave propagation through random media}, volume~4 of {\em
  Principles of statistical radiophysics}.
\newblock Springer-Verlag, Berlin; New York, 1989.

\bibitem{VAN_ROSSUM:1998}
M.~C.~W. van Rossum and T.~M. Nieuwenhuizen.
\newblock Multiple scattering of classical waves: from microscopy to mesoscopy
  and diffusion.
\newblock {\em Rev. Mod. Phys.}, 71(1):313--371, 1998.

\bibitem{Tsang_Advanced}
L.~Tsang and J.~A. Kong.
\newblock {\em Scattering of electromagnetic waves: advanced topics}.
\newblock Wiley, New York, 2001.

\bibitem{LAGENDIJK:1996}
A.~Lagendijk and B.~A. van Tiggelen.
\newblock Resonant multiple scattering of light.
\newblock {\em Phys. Rep.}, 270:143--215, 1996.

\bibitem{MANDEL:1995}
L.~Mandel and E.~Wolf.
\newblock {\em Optical coherence and quantum optics}.
\newblock Cambridge University Press, New York, 1995.

\bibitem{MANDEL:1965}
L.~Mandel and E.~Wolf.
\newblock Coherence properties of optical fields.
\newblock {\em Rev. Mod. Phys.}, 37(2):231--287, 1965.

\bibitem{FainIrreversibility}
B.~Fain.
\newblock {\em Irreversibilities in Quantum mechanics}.
\newblock Kluwer Academic Publishers, New York, 2002.

\bibitem{Riesz_FA}
F.~Riesz and B.~Sz\"okefalvi-Nagy.
\newblock {\em Functional Analysis}.
\newblock Dover, Mineola, 1990.

\bibitem{WOLF:1982}
E.~Wolf.
\newblock New theory of partial coherence in the space-frequency domain. {P}art
  {I}: spectra and cross spectra of steady-state sources.
\newblock {\em J. Opt. Soc. Am.}, 72(3):343--351, 1982.

\bibitem{BARABANENKOV:1991}
Y.~N. Barabanenkov and V.~D. Ozrin.
\newblock Asymptotic soution of the bethe-salpeter equation and the green-kubo
  formula for the diffusion constant for wave propagation in random media.
\newblock {\em Phys. Lett. A}, 152(1-2):38--42, 1991.

\bibitem{APALKOV:2004}
V.~M. Apalkov, M.~E. Raikh, and B.~Shapiro.
\newblock Incomplete photonic band gap as inferred from the speckle pattern of
  scattered light waves.
\newblock {\em Phys. Rev. Lett.}, 92(25):253902--1--253902--4, 2004.

\bibitem{VAN_TIGGELEN:1992}
B.~A. van Tiggelen, A.~Lagendijk, M.~P. van Albada, and A.~Tip.
\newblock Speed of light in random media.
\newblock {\em Phys. Rev. B}, 45(21):12233--12243, 1992.

\bibitem{SOUKOULIS:1994}
C.~M. Soukoulis, S.~Datta, and E.~N. Economou.
\newblock Propagation of classical waves in random media.
\newblock {\em Phys. Rev. B}, 49(6):3800--3811, 1994.

\bibitem{LIVDAN:1996}
D.~Livdan and A.~A. Lisyansky.
\newblock Transport properties of waves in absorbing random media with
  microstructure.
\newblock {\em Phys. Rev. B}, 53(22):14843--14848, 1996.

\bibitem{SAKODA:2005}
K.~Sakoda.
\newblock {\em Optical properties of photonic crystals}, volume~80 of {\em
  Optical Sciences}.
\newblock Springer, 2nd edition, 2005.

\bibitem{KUO:2004}
C.-H. Kuo and Z.~Ye.
\newblock Electromagnetic energy and energy flows in photonic crystals made of
  arrays of parallel dielectric cylinders.
\newblock {\em Phys. Rev. E}, 70:046617--046617--7, 2004.

\bibitem{ISHIMARU:1978}
A.~Ishimaru.
\newblock {\em Wave Propagation and Scattering in Random Media. Vol. 1. Single
  scattering and transport theory}.
\newblock Academic Press, New York, 1978.

\bibitem{VOLLHARDT:1980}
D.~Vollhardt and P.~Wolfle.
\newblock Diagrammatic, self-consistent treatment of the anderson localization
  problem in $d\leq 2$ dimensions.
\newblock {\em Phys. Rev. B}, 22(10):4666--4679, 1980.

\bibitem{Bir_Pikus_Symmetries}
G.~L. Bir and G.~E. Pikus.
\newblock {\em Symmetry and Strain-Induced Effects in Semiconductors}.
\newblock Wiley, New York, 1974.

\bibitem{VAN:1992:ID3086}
B.A. van Tiggelen, A.~Lagendijk, M.P. van Albada, and A.~Tip.
\newblock Speed of light in random media.
\newblock {\em Phys. Rev. B}, 45(21):12233--12243, 1992.

\bibitem{BARABANENKOV:1992:ID3303}
Y.N. Barabanenkov and V.D. Ozrin.
\newblock Problem of light diffusion in strongly scattering media.
\newblock {\em Phys. Rev. Lett.}, 69(6):1364--1366, 1992.

\bibitem{VAN:1993:ID3301}
B.A. van Tiggelen, A.~Lagendijk, and A.~Tip.
\newblock Comment on ``problem of light diffusion in strongly scattering
  media".
\newblock {\em Phys. Rev. Lett.}, 71(8):1284--1284, 1993.

\bibitem{JOHN:1996}
S.~John, G.~Pang, and Y.~Yang.
\newblock Optical coherence propagation and imaging in a multiple scattering
  medium.
\newblock {\em J. Biomed. Opt.}, 1(2):180--191, 1996.

\end{thebibliography}

\end{document}